\documentclass{aa}
\usepackage{graphicx}
\usepackage{txfonts}
\newcommand{\RSTAR}{\mbox{$R_{\star}$}}

\newcommand{\LSOL}{\mbox{$L_{\sun}$}}
\newcommand{\MSOLPERYR}{\mbox{$M_{\sun}$~yr$^{-1}$}}
\newcommand{\micron}{\mbox{$\mu$m}}
\newcommand{\KMS}{\mbox{km s$^{-1}$}}
\newcommand{\HOH}{\mbox{H$_2$O}}
\newcommand{\PERSQCM}{\mbox{cm$^{-2}$}}

\begin{document}
\title{
Imaging the dynamical atmosphere of 
the red supergiant Betelgeuse in the CO first overtone lines with VLTI/AMBER
\thanks{
Based on AMBER observations made with the Very Large Telescope 
Interferometer of 
the European Southern Observatory. Program ID: 082.D-0280 
(AMBER Guaranteed Time Observation)}
}

\author{K.~Ohnaka\inst{1} 
\and
G.~Weigelt\inst{1} 
\and
F.~Millour\inst{1,2}
\and
K.-H.~Hofmann\inst{1} 
\and
T.~Driebe\inst{1,3} 
\and
D.~Schertl\inst{1}
\and
A.~Chelli\inst{4}
\and
F.~Massi\inst{5}
\and
R.~Petrov\inst{2}
\and
Ph.~Stee\inst{2}
}

\offprints{K.~Ohnaka}

\institute{
Max-Planck-Institut f\"{u}r Radioastronomie, 
Auf dem H\"{u}gel 69, 53121 Bonn, Germany\\
\email{kohnaka@mpifr.de}
\and
Observatoire de la C\^{o}te d'Azur, Departement FIZEAU, 
Boulevard de l'Observatoire, B.P. 4229, 06304 Nice Cedex 4, 
France
\and
Deutsches Zentrum f\"ur Luft- und Raumfahrt e.V., 
K\"onigswinterer Str. 522-524, 53227 Bonn, Germany
\and
Institut de Plan\'etologie et d'Astrophysique de Grenoble, BP 53, 
38041 Grenoble, C\'edex 9, France
\and
INAF-Osservatorio Astrofisico di Arcetri, Instituto Nazionale di 
Astrofisica, Largo E. Fermi 5, 50125 Firenze, Italy
}

\date{Received / Accepted }

\abstract
{}
{
We present one-dimensional aperture synthesis imaging of the red 
supergiant Betelgeuse ($\alpha$~Ori) with VLTI/AMBER.  
We reconstructed for the first time one-dimensional images in the individual 
CO first overtone lines.  
Our aim is to probe the dynamics of the inhomogeneous atmosphere 
and its time variation. 
}
{
Betelgeuse was observed between 2.28 and 
2.31~\micron\ with VLTI/AMBER using the 16-32-48~m telescope 
configuration with a spectral resolution up to 12000 and an 
angular resolution of 9.8~mas.  
The good nearly one-dimensional $uv$ coverage allows us to reconstruct 
one-dimensional projection images (i.e., one-dimensional projections of the 
object's two-dimensional intensity distributions).  
}
{
The reconstructed one-dimensional projection images 
reveal that the star appears differently in the blue wing, 
line center, and red wing of the individual CO lines.  The one-dimensional 
projection 
images in the blue wing and line center show a pronounced, 
{{\em asymmetrically}} 
extended component up to $\sim$1.3~\RSTAR, while those in the red 
wing do not show such a component.  
The observed one-dimensional projection images in the lines can be reasonably 
explained by a model 
in which the CO gas within a region more than half as large as the stellar 
size is moving slightly outward with 0--5~\KMS, while the gas in the remaining 
region is infalling fast with 20--30~\KMS.  
A comparison between the {{\em CO line AMBER data}} taken in 2008 and 2009 
shows a significant time variation in the dynamics of the 
CO line-forming region in the photosphere and the outer atmosphere.  
In contrast to the line data, 
the reconstructed one-dimensional projection images in the continuum show only 
a slight deviation from a uniform disk or limb-darkened disk.  
We derive a uniform-disk diameter of $42.05 \pm 0.05$~mas and 
a power-law-type limb-darkened disk diameter of 
$42.49 \pm 0.06$~mas and a limb-darkening parameter of 
$(9.7 \pm 0.5) \times 10^{-2}$.  
This latter angular diameter leads to an effective temperature of 
$3690 \pm 54$~K for the continuum-forming layer. 
These diameters confirm that the near-IR size of Betelgeuse 
was nearly constant over the last 18 years, in marked contrast to the 
recently reported noticeable decrease in the mid-IR size. 
The {{\em continuum data}} taken in 2008 and 2009 reveal 
no or only marginal time variations, 
much smaller than the maximum variation predicted by the current three-dimensional 
convection simulations. 
}
{
Our two-epoch AMBER observations show that the outer atmosphere extending 
to $\sim$1.3--1.4~\RSTAR\ is asymmetric and its dynamics is dominated by 
vigorous, inhomogeneous large-scale motions, whose overall nature changes 
drastically within one year.  
This is likely linked to the wind-driving mechanism in red supergiants. 
}


\titlerunning{Imaging the dynamical outer 
atmosphere of Betelgeuse in the CO first overtone lines}
\authorrunning{Ohnaka et al.}
\maketitle

\section{Introduction}
\label{sect_intro}

Red supergiants (RSGs) experience slow, intensive mass loss up to 
$10^{-4}$~\MSOLPERYR, which is very important for understanding 
the final fate of massive stars.  
For example, our poor understanding of the RSG mass loss makes it 
difficult to estimate the main-sequence mass range of the progenitors of the 
Type IIP supernovae, which are the most common type of core-collapse 
supernovae.  
The mass loss also plays a significant role in the chemical enrichment of 
galaxies.  
Despite this importance, there are no satisfactory theories for the 
RSG mass loss at the moment, as stressed by Harper (\cite{harper10}).  

Studies of the outer atmosphere, where the winds are accelerated, 
are a key to tackling this problem.  
The outer atmosphere of RSGs has complicated structures.  
The UV observations of the well-studied RSG \object{Betelgeuse} 
($\alpha$~Ori, M1-2Ia-Ibe) with the Hubble Space
Telescope reveal that the hot ($\sim$6000--8000~K) chromospheric 
plasma is more than twice as extended as the photosphere 
(Gilliland \& Dupree \cite{gilliland96}).   
However, radio continuum observations with 
the Very Large Array show that much cooler ($\sim$1000--3000~K) 
gas extends to several stellar radii (Lim et al. \cite{lim98}), 
suggesting that the hot chromospheric plasma and cooler gas 
coexist.  
IR spectroscopic and interferometric studies of a few bright RSGs also show 
the presence of dense \HOH\ gas in the outer atmosphere, 
the so-called ``MOLsphere'', extending to 
$\sim$1.3--2.0~\RSTAR\ with column densities on the 
order of $10^{20}$~\PERSQCM\ and temperatures of 1500--2000~K 
(e.g., Tsuji \cite{tsuji00a}, \cite{tsuji00b}, \cite{tsuji06}; 
Ohnaka \cite{ohnaka04}; Perrin et al. \cite{perrin04}, \cite{perrin07}).  
The nature of the cool gas in the outer atmosphere of Betelgeuse 
has also recently been probed with mid-IR [Fe II] emission 
lines (Harper et al. \cite{harper09a}).  
Possibly the chromospheric plasma with a small filling factor 
is embedded in more abundant, cooler gas. 
This inhomogeneous, multi-component nature of the outer atmosphere is 
considered to play a crucial role in driving mass outflows in 
RSGs.  
There is observational evidence for the asymmetric, inhomogeneous nature 
of the circumstellar material. 
For Betelgeuse, Kervella et al. (\cite{kervella09}) found a very faint plume 
extending to $\sim$6~\RSTAR\ in the near-IR, 
while a millimeter CO map shows a blob at 
$\sim$5\arcsec\ ($\sim$235~\RSTAR) away from the star 
(Harper et al. \cite{harper09b}).  
The emission of the CO fundamental lines near 4.6~\micron\ shows that the 
circumstellar envelope is approximately spherical within 
3\arcsec\ (140~\RSTAR) but with signatures of mildly clumpy structures 
(Smith et al. \cite{smith09}).

High-spectral and high-spatial resolution observations of strong IR molecular 
lines are ideal for probing the physical properties of the 
outer atmosphere.  
The near-IR interferometric instrument AMBER (Astronomical Multi-BEam 
combineR) at the Very Large Telescope Interferometer (VLTI) 
is well suited for this goal with its high spectral 
resolution up to 12000 and high spatial resolution down to 1--2~mas with 
the current maximum baseline of 130~m.  
In 2008, we observed Betelgeuse in the CO first overtone lines near 
2.3~\micron\ 
with AMBER (Ohnaka et al. \cite{ohnaka09}, 
hereafter Paper I).  
The high spectral resolution of AMBER allowed us to 
detect salient signatures of inhomogeneities in the individual CO lines and to 
spatially resolve the gas motions in a stellar photosphere (and 
also MOLsphere) for the first time other than the Sun. 

However, in 2008, we obtained data only at six $uv$ points, which are 
insufficient for image reconstruction. 
In order to obtain a more complete picture of the dynamics of the inhomogeneous 
outer atmosphere, we carried out new AMBER observations of Betelgeuse 
with a better $uv$ coverage in 2009.  
In this second paper, we report on the first one-dimensional aperture synthesis 
imaging of Betelgeuse in the CO first overtone lines, 
as well as on time variation in the dynamics of the stellar atmosphere 
in an interval of one year. 
The paper is structured as follows.  The AMBER observations, data reduction, 
and image reconstruction are outlined in Sect.~\ref{sect_obs}.  
We describe the results about the time variations as well as the 
one-dimensional image reconstruction in Sect.~\ref{sect_res}.  
The modeling of the velocity field 
presented in Sect.~\ref{sect_modeling} is followed by the discussion on 
the dynamics in the extended outer atmosphere (Sect.~\ref{sect_discuss}).  
Conclusions are presented in Sect.~\ref{sect_conclusions}.

\section{Observations}
\label{sect_obs}

\subsection{AMBER observations and data reduction}
\label{subsect_obs_red}

AMBER (Petrov et al. \cite{petrov07}) is the near-IR (1.3---2.4~\micron) 
spectro-interferometric instrument at VLTI, which combines three 8.2~m Unit 
Telescopes (UTs) or 1.8~m Auxiliary Telescopes (ATs). 
AMBER measures the amplitude of the Fourier transform---the so-called 
visibility or visibility amplitude and two observables that 
contain information about the phase of the object's Fourier transform: 
differential phase (DP) and closure phase (CP).  
The DP roughly represents how the object's phase in a spectral feature deviates 
from that in the continuum.  Non-zero DP represents information about the 
photocenter shift in a spectral feature with respect to the 
continuum. 
The CP is the sum of the measured Fourier phases around a closed triangle of 
baselines (i.e., $\varphi_{12} + \varphi_{23} + \varphi_{31}$), not affected 
by the atmospheric turbulence.  
The CP is always zero or $\pi$ for point-symmetric objects, and 
non-zero and non-$\pi$ CPs indicate an asymmetry in the object. 
Moreover the CP is important for aperture synthesis imaging in optical/IR 
interferometry. 

Betelgeuse was observed on 2009 January 5 and 6 with AMBER using three ATs 
in the E0-G0-H0 linear array configuration 
with 16--32--48~m baselines (AMBER Guaranteed Time Observation, 
Program ID: 082.D-0280, P.I.: K.~Ohnaka).  
As in Paper I, 
we used the $K$-band high-resolution mode (HR\_K) with a spectral 
resolution of 12000 covering wavelengths from 2.28 to 2.31~\micron\ to 
observe the strong CO first overtone lines 
near the (2,0) band head.  
Fringes could be 
detected on all three baselines without the VLTI fringe tracker FINITO.  
We obtained a total of 54 data sets on two half nights.  
The data sets taken more than $\sim$2 minutes apart were treated as 
separate data sets, because Betelgeuse is 
strongly resolved and the visibility varies noticeably even for a slight 
change in the baseline length (see Fig.~\ref{vis_continuum}).  
As shown in Fig.~\ref{uvcoverage}, 
the observed $uv$ points align 
approximately linearly at position angles of $73 \pm 2$\degr.  
This linear $uv$ coverage allows us to 
sample the visibility function densely at this particular position angle 
and to reconstruct one-dimensional projection images 
as described in Sect.~\ref{subsect_imaging}.  
Each data set consists of 500 frames (NDIT) with each frame taken with 
a detector integration time (DIT) of 120~ms.  
\object{Sirius} (\object{$\alpha$~CMa}, A1V, $K$ = $-1.4$) was observed 
for the calibration of the interferometric data of Betelgeuse.  
We adopted the same angular diameter of $5.6 \pm 0.15$~mas from 
Richichi \& Percheron (\cite{richichi05}) as adopted in Paper I.  
We only used the calibrator data sets obtained just before and after 
each data set on Betelgeuse.  
A summary of the observations is given in Table~\ref{obslog}.

We reduced our AMBER data with amdlib ver.2.2\footnote{
Available at http://www.jmmc.fr/data\_processing\_amber.htm}, 
which is based on the P2VM algorithm (Tatulli et al. \cite{tatulli07}).  
Some data sets, particularly those measured on the longest baseline 
and/or near the CO band head, are too noisy for the analysis.  
Therefore, we improved the SNR by binning the entire raw data (object, dark, 
sky, and P2VM calibration data) in the spectral direction with a 
running box car function as described in Paper I.  
We used different binnings with these spectral resolutions: 

\begin{enumerate}
\item Spectral resolution = 12000 (no binning), 8000, and 4800 
  for the data sets \#16, \#17, \#18, and \#45--\#49 for 
  comparison with the 2008 data (Sect.~\ref{subsect_res_cont} and 
  \ref{subsect_res_CO}). 
\item Spectral resolution = 6000 for all data sets for
  the image reconstruction in the continuum and in the {\em individual, 
    isolated CO lines} (Sect.~\ref{subsect_img_cont} and 
~\ref{subsect_img_lines}, Fig.~\ref{image_lines}). 
\item Spectral resolution = 1600 for all data sets for
  the image reconstruction near the {\em CO band head} at 2.294~\micron\ 
  (Sect.~\ref{subsect_img_bh}, Fig.~\ref{image_bh}). 
\end{enumerate}

We checked for a systematic difference in the calibrated visibilities 
and differential/closure phases by taking 
the best 20\%, 50\%, and 80\% of all frames in terms of the fringe SNR 
(Tatulli et al. \cite{tatulli07}). 
The difference between the results obtained with the best 20\% and 80\% 
frames is typically $\sim$10\%. 
We took the best 20\% of all frames for our final visibilities to 
avoid the systematic bias due to the rapid atmosphere and to keep 
decent SNRs in the final results.  
On the other hand, the calibrated differential/closure phases do not show 
this systematic dependence on the frame selection criterion.  
Therefore, we included the best 80\% of all frames for the final DPs and 
CPs.  
The errors of the resulting visibilities, DPs, and CPs were estimated 
in the same manner as in Paper I.

We removed telluric lines from the observed spectra of Betelgeuse 
as best as possible by using Sirius as a spectroscopic standard star.  
The telluric lines identified in the spectrum of Sirius were also used 
for wavelength calibration.  As a template of the telluric lines, 
we convolved the atmospheric transmission spectra measured at the Kitt Peak 
National Observatory\footnote{http://www.eso.org/sci/facilities/paranal/instruments/isaac/tools/\\spectra/atmos\_S\_K.fits} 
to match the spectral resolutions of the data.  
The uncertainty in wavelength calibration is 
$2.0 \times 10^{-5}$~\micron\ (2.6~\KMS).  
We note that the uncertainties in wavelength calibration for the 
2008 data and 2006 data in Paper~I were mistakenly overestimated.  
The correct uncertainties in the 2008 data and 2006 data 
are $2.3 \times 10^{-5}$~\micron\ (3.0~\KMS) and $1.8 \times 10^{-4}$~\micron\
(25.1~\KMS), respectively.

\subsection{One-dimensional image reconstruction}
\label{subsect_imaging}

The good linear $uv$ coverage along the position angle of 73\degr\ 
shown in Fig.~\ref{uvcoverage} provides an opportunity to reconstruct 
the so-called one-dimensional projection image, 
which is obtained by 
integrating the object's two-dimensional intensity distribution along the direction 
perpendicular to the linear $uv$ coverage on the sky 
(central slice theorem or Fourier slice theorem).  
In other words, this one-dimensional projection image represents 
the two-dimensional intensity distribution compressed or squashed onto the 
linear $uv$ coverage on the sky. 
For example, the one-dimensional projection image of a uniform disk 
is a semi-circle (see also the two-dimensional image of a limb-darkened disk 
and its one-dimensional projection image shown in Figs.~\ref{simdata_ldd}a and 
\ref{simdata_ldd}c). 
The reconstruction of one-dimensional projection images was first proposed for 
radio interferometry by Bracewell (\cite{bracewell56}).  
Whereas the information in the direction perpendicular to the baseline 
vector is lost in one-dimensional projection images, they still provide model-independent 
information about the geometrical extent and asymmetry of the object.  
The reconstruction of one-dimensional projection images 
from IR interferometric data or lunar occultation data has 
been carried out (e.g., Navarro et al. \cite{navarro90}; 
Leinert et al. \cite{leinert91}; Tatebe et al. \cite{tatebe06}; 
Chandler et al. \cite{chandler07}).

We used the MiRA package 
ver.~0.9.9\footnote{http://www-obs.univ-lyon1.fr/labo/perso/eric.thiebaut/mira.html} 
(Thi\'ebaut et al. \cite{thiebaut08}) to reconstruct one-dimensional projection images 
at each spectral channel (details of our image reconstruction procedure are 
described in Appendix~\ref{appendix_simtests}).  
We first carried out the image reconstruction using computer-simulated 
data to examine effects of the $uv$ coverage and reconstruction parameters 
such as the initial model, prior, and regularization scheme on the 
reconstructed images. 
These tests with simulated data are crucial for examining the credibility 
of aperture synthesis imaging particularly for objects with complex 
structures.  

With appropriate reconstruction parameters determined from these tests, 
we attempted to reconstruct one-dimensional projection images from the observed 
162 visibility amplitudes and 54 CPs.  
While this worked well for the continuum, 
the reconstruction in the CO lines turned out to be very sensitive to the 
reconstruction parameters.  
For example, depending on the size of the uniform disk used as the 
initial model, the reconstructed one-dimensional projection image in the CO lines 
shows a faint region on the eastern or western side.  
Therefore, we used the self-calibration technique, 
which has recently been successfully applied to AMBER data 
for the first time by Millour et al. (\cite{millour11}).  We added 
modifications to their technique to deal with some issue specific 
to our data of Betelgeuse as described in Appendix \ref{appendix_selfcal}. 
This technique allows us to restore the phase of the complex Fourier 
transform of the object's intensity distribution from the DP 
measurements.  
Image reconstruction with the complex visibility (i.e., visibility amplitude 
and phase) removes the ambiguity of the solution derived with the 
visibility amplitude and CP alone.

\section{Results}
\label{sect_res}

\subsection{Continuum data}
\label{subsect_res_cont}

To compare with the $K$-band continuum visibilities from the 2008 data that 
were derived from the binned data with a spectral resolution 4800, 
we also derived the visibilities from the 2009 data binned with the 
same spectral resolution.  
As in Paper I, we selected continuum points shortward of the CO band head 
at 2.294~\micron.  
For each data set, we averaged the visibilities over the selected continuum 
points.  We took the simple mean of the errors as the errors in the average 
continuum visibilities without 
reducing by $\sqrt{N_{\rm cont}} $, where $N_{\rm cont}$ is the number of 
the selected continuum points.  The reason is that the measurement errors 
are dominated by the systematic error in the absolute 
visibility calibration and do not become smaller by the averaging.  
We applied this averaging to the 2008 data as well.  
Since the different continuum spectral channels correspond to 
slightly different spatial frequencies, we also averaged 
the spatial frequencies from the selected continuum points.

\begin{figure}
\resizebox{\hsize}{!}{\rotatebox{0}{\includegraphics{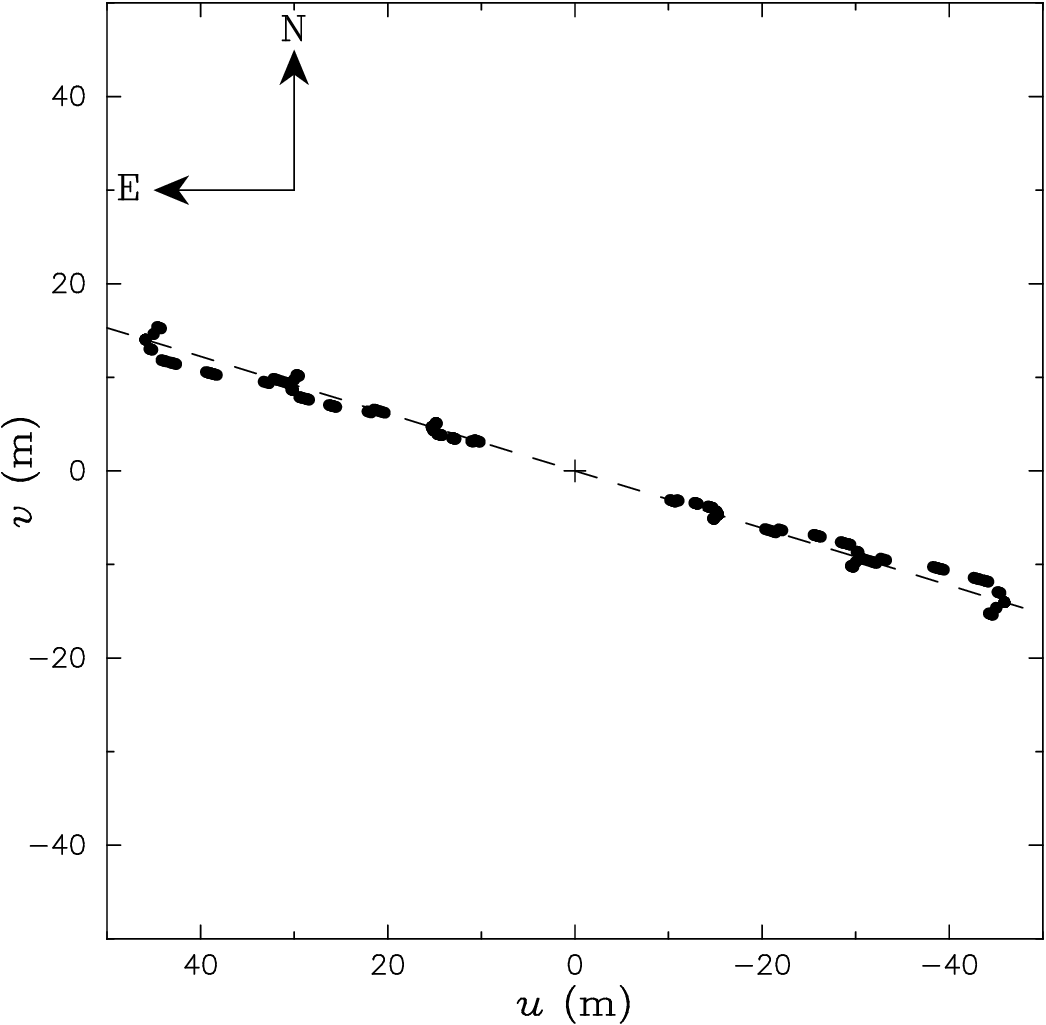}}}
\caption{
$uv$ coverage of our AMBER observations of Betelgeuse.  The dashed line 
represents the average position angle of 73\degr. 
}
\label{uvcoverage}
\end{figure}

\begin{figure*}
\resizebox{\hsize}{!}{\rotatebox{-90}{\includegraphics{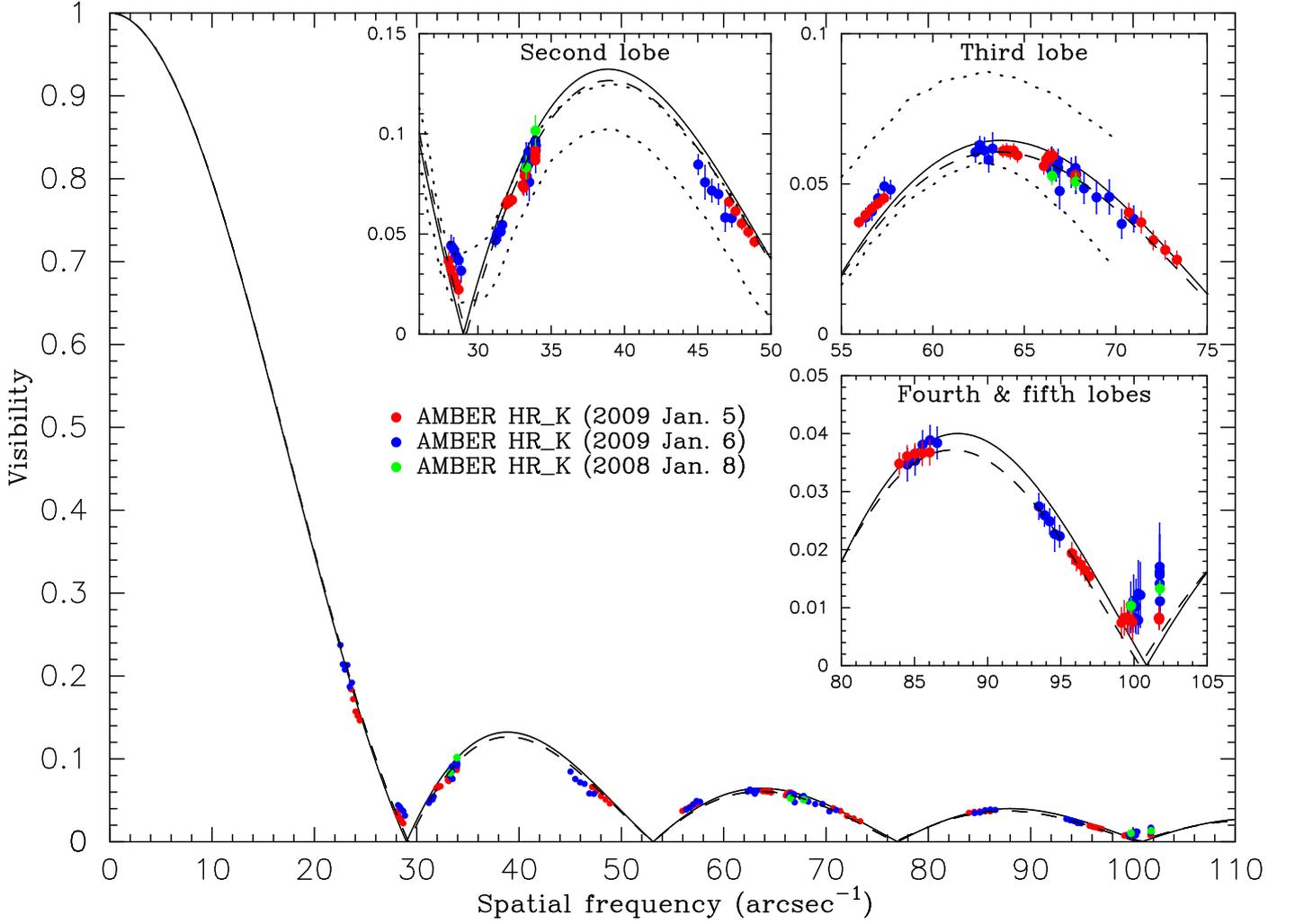}}}
\caption{Continuum visibilities of Betelgeuse averaged over the continuum 
spectral channels between 2.28 and 2.293~\micron.  
The insets show enlarged views of the second, third, 
and fourth/fifth lobes.  
The solid and dashed lines represent the visibilities for a uniform disk with 
a diameter of 42.05~mas and for a limb-darkened disk with a diameter of 
42.49~mas 
and a limb-darkening parameter of 0.097 (power-law-type limb-darkened disk 
of Hestroffer \cite{hestroffer97}), respectively.  
The dotted lines represent the maximum range of the 
variations in the 2.22~\micron\ visibility due to time-dependent 
inhomogeneous surface structures predicted by the three-dimensional convection simulation 
of Chiavassa et al. (\cite{chiavassa09}), who presents the 
model prediction up to 70 arcsec$^{-1}$.  
}
\label{vis_continuum}
\end{figure*}

Figure~\ref{vis_continuum} shows the $K$-band continuum visibilities measured 
in 2008 and 2009 as a function of spatial frequency.  
The figure reveals that the nearly linear $uv$ coverage shown in 
Fig.~\ref{uvcoverage} 
enabled us to sample the visibility function 
quite densely from the first to the fifth visibility lobe.  
Uniform-disk fitting to the 2009 data results in a diameter 
of $42.05 \pm 0.05$~mas with a reduced $\chi^2$ of 3.8.  
Fitting with a power-law-type limb-darkend disk (Hestroffer et 
al. \cite{hestroffer97}) results in a limb-darkened disk diameter of 
$42.49 \pm 0.06$~mas and a limb-darkening parameter of $(9.7 \pm 0.5) \times 
10^{-2}$ with a better reduced $\chi^2$ of 2.5.  
While the reduced $\chi^2$ value 
is still higher than 1, Fig.~\ref{vis_continuum} shows that the deviation from 
the limb-darkened disk is not strong, as found for the 2008 data.  
Only at the highest spatial frequency (i.e., the smallest spatial scale) 
is the deviation noticeable, but the errors are also large there.

The limb-darkened disk diameter derived from the $K$-band continuum data 
and a bolometric flux of $(111.67 \pm 6.49) \times 10^{-13}$~W~cm$^{-2}$ 
(Perrin et al. \cite{perrin04}) lead to an effective temperature of 
$3690 \pm 54$~K.  We propose this value as an effective temperature of 
the continuum-forming layer, approximately free from the effects of molecular 
lines.  
Perrin et al. (\cite{perrin04}) modeled $K$-broadband interferometric 
measurements of Betelgeuse with a continuum-forming blackbody sphere 
and an extended molecular shell.  They derived $3690 \pm 50$~K for the 
continuum-forming sphere.  This value excellently agrees with our effective 
temperature of the continuum-forming layer.  
Our effective temperature is slightly higher than the $3600 \pm 66$~K 
recently derived by Haubois et al. (\cite{haubois09}) from the the $H$-band 
observations with the Infrared Optical Telescope Array (IOTA), but 
both agree within the uncertainties.  

Figure~\ref{vis_continuum} reveals that the continuum visibilities show 
no or only marginal time variations between 2008 (green dots) and 2009 (red 
and blue dots) within the measurement errors.  
We compare this observational result with the current three-dimensional 
convection simulation for RSGs by Chiavassa et al. (\cite{chiavassa09}).  
The visibility predicted for 2.2~\micron\ (Fig.~18 of Chiavassa et al. 
\cite{chiavassa09}), which approximately samples the continuum, shows 
maximum time variations of, for example, $\pm$40\% in the third lobe. 
However, the visibilities observed in the third lobe 
(at spatial frequencies of 66--68~arcsec$^{-1}$) show no time 
variation within the error bars (3--5\%), 
and these error bars are 8--13 times smaller than the predicted maximum 
variation.  
While it is not very likely that we observed Betelgeuse at two 
epochs when it accidentally showed the same, weak deviations from the 
limb-darkened disk, this possibility cannot be entirely excluded.  
However, 
it is also possible that the current three-dimensional convection simulation predicts 
too pronounced surface structures and time variations 
owing to the gray approximation adopted for the radiative transfer, 
as Chiavassa et al. (\cite{chiavassa09}) mention.  
The most direct test for three-dimensional convection simulations 
is to measure the amplitude of the temporal 
fluctuations in the visibility and closure phase as well as the time scale 
of fluctuations by long-term monitoring observations and compare these 
with the model predictions.  
AMBER observations at more epochs would be necessary to 
draw a definitive conclusion about whether or not Betelgeuse 
seen in the continuum indeed 
shows much weaker inhomogeneities and much smaller time variations 
than predicted by the current three-dimensional convection simulation.  

The deviation from the limb-darkened disk in the $K$-band continuum 
visibilities is lower than that observed in the $H$ broadband by 
Haubois et al. (\cite{haubois09}).  
Their measurements show deviations of the visibilities from the limb-darkened 
disk as high as 80--120\% already in the fourth lobe 
(converted from the squared visibilities plotted in their 
Fig. 4), where our $K$-band continuum data still follow the 
limb-darkened disk within the measurement errors of 5--10\% 
except for the data points near the visibility null at 
$\sim$100~arcsec$^{-1}$. 
The cause of this difference is not yet clear, because of 
a number of differences between their observations and ours 
(e.g., differences in the observed wavelengths, spectral resolution, 
position angle coverage).  
$H$-band observations with higher spectral resolution and/or 
$K$-band observations with a wider position angle coverage
are necessary to clarify this issue.

\subsection{Long-term behavior of the near-IR and mid-IR angular diameters}
\begin{figure}
\resizebox{\hsize}{!}{\rotatebox{-90}{\includegraphics{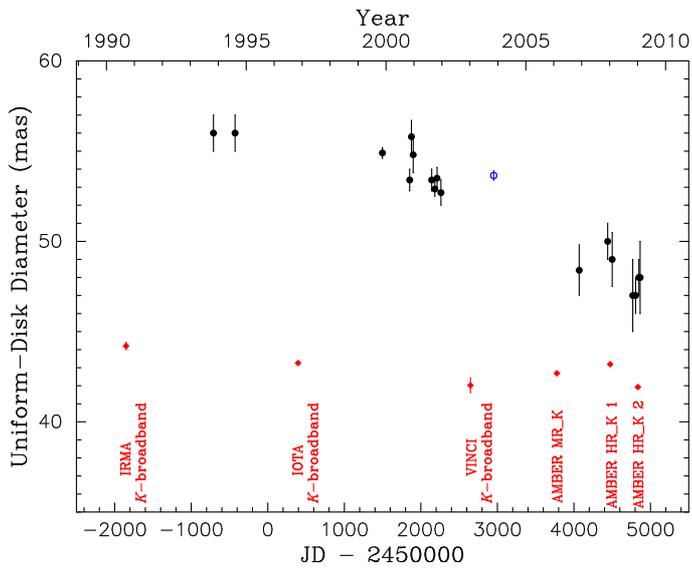}}}
\caption{
Uniform-disk diameters of Betelgeuse measured in the $K$ band and at 
11~\micron\ as a function of time.  
The black filled circles and the blue open circle represent the 11~\micron\ 
diameters measured by Townes et al. (\cite{townes09}) and Perrin et al. 
(\cite{perrin07}), respectively.  
The diamonds represent the $K$-band diameters from the following 
references. 
IRMA: Dyck et al. (\cite{dyck92}). 
IOTA: Perrin et al. (\cite{perrin04}). 
VINCI: Paper I. 
AMBER MR\_K: AMBER medium-resolution ($\lambda / \Delta \lambda$ = 1500)
data from Paper I. 
AMBER HR\_K 1: AMBER high-resolution data from Paper I. 
AMBER HR\_K 2: Present paper. 
The errors in the IOTA and AMBER measurements are smaller than the symbols. 
}
\label{alfori_UD_time}
\end{figure}

Figure~\ref{alfori_UD_time} shows the $K$-band uniform-disk diameters 
of Betelgeuse 
from the literature and the archival data summerized in Paper I, together 
with the 11~\micron\ uniform-disk diameter presented in Townes et al. 
(\cite{townes09}), who found a noticeable decrease in the 
11~\micron\ size in the last 15 years, 
and the one-epoch measurement by Perrin et al. (\cite{perrin07}).  
In marked contrast to the noticeable decrease in the 11~\micron\ size, 
the $K$-band diameter has been quite stable for the last 18 years 
with only a possible, slight long-term decrease.  
These results can be qualitatively explained as follows. 
While the size of the star itself has been stable over the last 18 years, 
the temperature and densities and/or shape of the outer atmosphere 
have changed significantly (e.g., decrease in temperature and/or density). 
Because the mid-infrared apparent size is largely affected by the MOLsphere 
and by dust (Ohnaka \cite{ohnaka04}; Verhoelst et al. \cite{verhoelst06}; 
Perrin et al. \cite{perrin07}), the changes in the outer atmosphere 
lead to a noticeable change in the 11~\micron\ size.  
This interpretation has also been recently reached by Ravi et al. 
(\cite{ravi10}) based on the estimation of the surface temperature 
seen at 11~\micron.  
On the other hand, the angular size measured with the $K$-broadband filter 
is only slightly affected, because the strong molecular bands of CO 
and \HOH\ are present only at the short and long wavelength edge 
of the $K$ band.  
Detailed modeling of the mid-IR interferometric and spectroscopic 
data, which is necessary to quantitatively derive the change in the physical 
properties of the MOLsphere and dust, is beyond the scope of this paper 
and will be pursued in a forthcoming paper.

\subsection{Significant time variation in the CO first overtone lines}
\label{subsect_res_CO}

\begin{figure*}
\resizebox{\hsize}{!}{\rotatebox{0}{\includegraphics{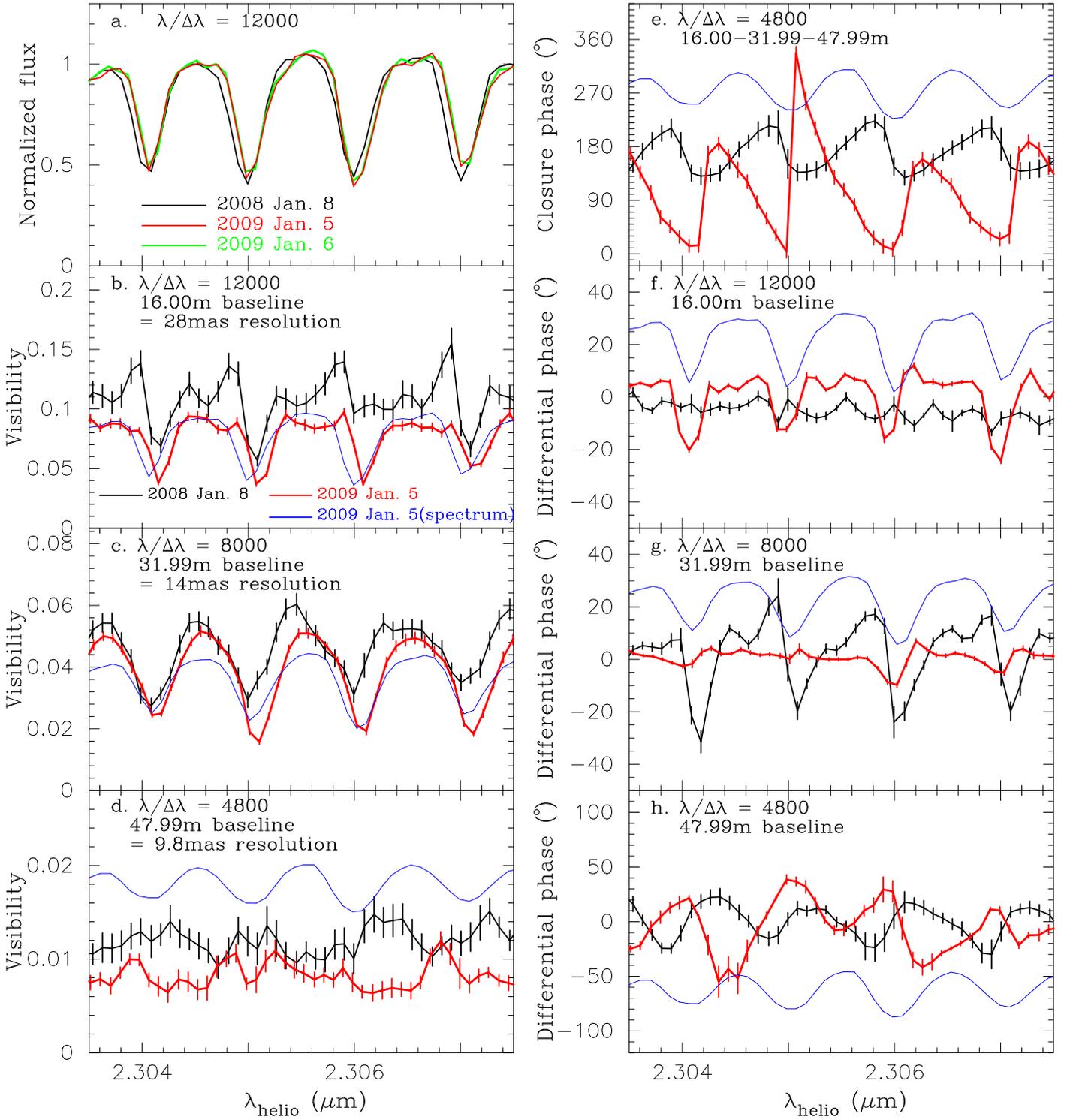}}}
\caption{
AMBER data in the CO first overtone lines obtained in 2008 and 2009.  
The black and red solid lines represent the data taken in 2008 and 2009, 
respectively.  In the panels {\bf b--h}, the scaled spectrum observed on 
2009 Jan 5 is overplotted 
by the blue solid line to show the asymmetry of the visibilities and 
differential/closure phases within each line profile.  
The spectral resolutions given in the panels are the values of the 
binned data as described in Sect.~\ref{subsect_obs_red}.   
{\bf a:} Normalized spectra.  Two spectra derived from the data taken on 
2009 Jan 5 and 6 are plotted by the red and green solid lines, respectively, 
while the 2008 spectrum is plotted by the black solid line. 
{\bf b--d:} Visibilities observed on the E0-G0-16m, G0-H0-32m, and E0-H0-48m 
baselines.  
{\bf e:} Closure phase. 
{\bf f--h:} Differential phases observed on the E0-G0-16m, G0-H0-32m, and 
E0-H0-48m baselines.  
}
\label{obsresCO}
\end{figure*}

Figure~\ref{obsresCO} shows a comparison between the data taken in 2008 and 
2009 for four representative CO lines.  
The results for the 2009 data were obtained from the merged data 
of the data sets \#16, \#17, and \#18, which were taken at $uv$ points 
very close to one of the data sets obtained in 2008 (data set \#1 in 
Paper I).  
The data sets \#45--\#49 were also taken at $uv$ points very close to 
the same 2008 data set, and the results from these data sets agree 
well with those shown in Fig.~\ref{obsresCO}.  
We used the binning with the same spectral resolution as applied to the 
data from 2008: 
spectral resolution of 12000 (no binning) for the 16~m baseline data, 
8000 for the 32~m baseline data, and 4800 for the 48~m baseline 
data and CP, respectively.

In marked contrast to the continuum data, Fig.~\ref{obsresCO} reveals 
significant time variations in the CO line visibilities.  
The visibility within each CO line obtained on the 16~m baseline in 2008 
was characterized by the maxima in the blue wing and minima in the red wing 
(black line in Fig.~\ref{obsresCO}b). 
In the 2009 data, the visibility on the 16m baseline does not show the 
maxima in the blue wing anymore (red line in Fig. \ref{obsresCO}b) 
and is characterized only by the minima in the red wing. 
The visibilities on the 32~m and 48~m baselines also show time variations, 
although the data on the 48~m baseline are noisy.  

Time variations are even clearer in the DPs and CPs. 
Non-zero DPs were not detected in the CO lines on the 16m baseline in 2008, 
but the 2009 data show clear non-zero DPs in the CO lines.  
On the other hand, the non-zero DPs on the 32m baseline obtained in 2009 
are much weaker than those in the 2008 data.  The DPs on the 48m baseline 
as well as the CPs measured in 2009 also show significant time variations.  
The non-zero DPs and non-zero/non-$\pi$ CPs indicate the asymmetry of the 
CO-line-forming region in 2009, as found in 2008.  

The observed spectra also reveal changes in the line profiles.  
The lines observed in 2009 are redshifted by $\sim$6~\KMS\ compared to 
those observed in 2008, as shown in Fig.~\ref{obsresCO}a.  
The spectra taken on 2009 Jan 5 and 6 agree very well, although they were 
calibrated independently.  This confirms that the redshift of the CO lines in 
the 2009 data is real.  
All these results suggest that the dynamics in the atmosphere of 
Betelgeuse has changed in an interval of one year.

\subsection{One-dimensional projection images in the continuum}
\label{subsect_img_cont}

The reconstructed one-dimensional projection image in the continuum at 
2.30662~\micron\ is shown by the black line in 
Fig.~\ref{image_lines} 
(comparison between the observed interferometric data and those from the 
reconstructed image is shown in Fig.~\ref{fit_selfcal}).  
Also plotted is the one-dimensional projection image of the limb-darkened disk 
with the angular diameter of 42.49~mas and the limb-darkening parameter 
of 0.097 (gray line, overlapping with the green line at angular 
distances between $+15$ and $-15$~mas).  
These one-dimensional projection images are already convolved with the 
Gaussian beam with a FWHM of 9.8~mas as described in 
Appendix~\ref{appendix_simtests}. 
Figure~\ref{image_lines} shows that the stellar surface is well resolved with 
the beam size of 1/4 of the diameter of the stellar disk.  
The one-dimensional projection images obtained at all continuum spectral 
channels 
agree well within the uncertainty of the reconstruction ($\sim$1\%). 

The overall deviation from the limb-darkened disk is small, $\sim$5\% 
on the eastern side (position angle = 73\degr).  
Because inhomogeneities in the direction perpendicular to the baseline 
vector on the sky are smeared out in the one-dimensional projection image, 
we estimated the upper limit on the strength of inhomogeneities using 
a uniform disk with one Gaussian-shaped dark spot. 
The $H$-band image of Betelgeuse shows spots with FWHMs up to 10--11~mas 
(Haubois et al. \cite{haubois09}), 
which we adopt for our model.  For a given amplitude of the spot,  
we generated 10000 models with random 
positions of the spot and 
counted the number of models whose one-dimensional projection image shows 
deviation from the uniform disk smaller than 5\%.  
For spot amplitudes lower than 20\% of the stellar disk intensity, 
more than 70\% of the models show deviations compatible with the 
observations. 
However, 
the fraction of these models is 13\%, 3\%, and $<$0.01\% for spot amplitudes of 
30\%, 40\%, and 50\% of the stellar disk intensity, respectively. 
Therefore, we estimate the amplitude of the spot to be smaller than 
20--30\% of the stellar disk intensity with flux contributions of 1.6--2.4\%.

\subsection{One-dimensional projection images in the individual CO lines}
\label{subsect_img_lines}

Figure~\ref{image_lines} shows the one-dimensional projection images reconstructed 
{\em in the blue wing, line center, and red wing 
within the CO line} centered at 2.3061~\micron\ because of the two 
transitions (2,0) $R(26)$ and $R(75)$ with a spectral resolution of 6000 
(comparison between the observed interferometric quantities and those 
from the reconstructed images is shown in Fig.~\ref{fit_selfcal}). 
The one-dimensional projection images are normalized with the peak intensities 
but are not artificially registered with one another, because the 
relative astrometry is preserved thanks to the restored visibility phase. 
The round or blunt shape of the images primarily results from the 
projection of the two-dimensional images onto the baseline vector on the sky. 
The level of the image reconstruction noise 
is estimated to be $\sim$1.5\% from the strengths of the artifacts in the 
entire 
field of view used for the reconstruction (256 mas).  
The one-dimensional projection images in other, isolated CO lines 
above $\sim$2.3~\micron\ agree well with those shown in 
Fig.~\ref{image_lines}, which adds fidelity to the image reconstruction.  
{\em These one-dimensional projection images represent the imaging of 
  the photosphere and MOLsphere of an RSG, for the first time, in the 
  individual CO first overtone lines.  }

Clearly, the one-dimensional projection images 
in the blue wing and line center are more extended than those 
in the continuum, with the extension of 15\% and 30\% on the 
eastern and western side, respectively, when measured at the noise 
level of the image reconstruction ($\sim$1.5\% of the peak 
intensities). 
The actual geometrical extension with intensities lower than the 
reconstruction noise level could be even larger.  
These values are 
roughly consistent with the size of the MOLsphere probed with the \HOH\ 
bands (e.g., Tsuji \cite{tsuji00a}, \cite{tsuji06}; 
Ohnaka \cite{ohnaka04}; Perrin et al. \cite{perrin04}, \cite{perrin07}).  
On the other hand, the red wing one-dimensional projection image 
shows only a slight deviation from the continuum without a trace of the 
extended component.  
Furthermore, 
the different appearance of the extended component in the blue and red wing 
suggests that the vigorous, inhomogeneous gas motions are present 
not only in the photosphere extending to $\sim$1.1 \RSTAR\ 
(see Sect.~4 in Paper I) but also 
in the layers extending to $\sim$1.3~\RSTAR.  This is because 
if the inhomogeneous gas motions were present only in the photosphere, the 
extended component would appear the same in the blue and red wing.

\subsection{One-dimensional projection images in the CO band head}
\label{subsect_img_bh}

Figure~\ref{image_bh} shows the one-dimensional projection images reconstructed 
{\em at four different wavelengths} in the CO (2,0) band head with a 
spectral resolution of 1600 
(see Fig.~\ref{fit_bh} for a comparison 
between the observed interferometric quantities and those from the 
reconstructed images).  
The peak of the one-dimensional projection image at the band head is shifted 
to the east 
with respect to that in the continuum and has 
geometrical extensions of 18\% and 13\% on the eastern and western side, 
respectively, when measured at the noise level of the image reconstruction.

In none of the reconstructed images in the continuum, CO lines, and 
CO band head did we detect a feature corresponding to the faint plume 
reported by Kervella et al. (\cite{kervella09}), 
although their (single-dish) observations 
were carried out coincidentally almost simultaneously with our AMBER 
measurements.  
This is presumably because of the very faint nature of the plume. 
Its intensity is below 1\% of the center of the stellar disk even at 
$\la$1.3~\micron, where it appears the most prominent, and 
the plume is even less pronounced at 2.12 and 2.17~\micron.  
Such a faint structure is below the noise of the image 
reconstruction from the present data (1.5\% and 3\% in 
Figs.~\ref{image_lines} and \ref{image_bh}, respectively). 
However, future observations with better accuracies 
could reveal the presence of the plume in the CO lines, which would 
be useful for understanding the nature of the plume.  
Furthermore, the reconstructed images in the CO band head with a 
spectral resolution 1600 can be compared with future medium-spectral 
resolution AMBER observations.

\section{Modeling of the velocity field}
\label{sect_modeling}

We used our stellar patch model presented in Paper I to characterize the 
velocity field in the photosphere and MOLsphere.  This model consists of two 
CO layers that represent the photosphere and MOLsphere.  An inhomogeneous 
velocity field is represented by a patch (or clump) of CO gas moving at some 
velocity different from the CO gas in the remaining region.  
Because the reconstructed one-dimensional projection images do not allow us to 
know the actual number and shape of the patches, we assumed only one patch in 
our modeling to keep the number of free parameters as small as possible.  
Furthermore, for the temperature, CO column density, and the radius of the two 
layers, we used the same parameters as derived in Paper I for the following 
reason.  
As discussed in Sect.~\ref{sect_res}, the CO line profiles observed in 2008 
and 2009 show little time variation in the line depth and width except for the 
redshift.  
This implies that the physical 
properties of the photosphere and MOLsphere, such as the density and
temperature, may not have changed significantly, although there must have 
been temporal and spatial fluctuations.  
The parameters adopted from Paper I are as follows: 
the inner CO layer is assumed to be 
located at 1.05 \RSTAR\ with a temperature of 2250~K and 
a CO column density of $5 \times 10^{22}$~\PERSQCM, while the outer CO layer 
is assumed to be located at 1.45~\RSTAR\ with a temperature of 1800 K and a 
CO column density of $1 \times 10^{20}$~\PERSQCM.  
We also adopted a microturbulent velocity of 5~\KMS\ for both layers as in 
Paper I.  
This means that we attempt to explain the one-dimensional projection images observed in 
2009 by changes in the velocity field, as well as in the position and size 
of the patch.  
The wavelength scale of the model spectra was converted to the 
heliocentric frame assuming a heliocentric velocity of 20~\KMS\ 
(Huggins \cite{huggins87}; Huggins et al. \cite{huggins94}).

Figure~\ref{alfori2_model} shows the two-dimensional images, one-dimensional projection images, 
and the spectrum predicted by the best-fit model for the same CO line as shown 
in Fig.~\ref{image_lines}. 
The model is characterized by a large, off-centered, circular patch of CO gas, 
which dominates the upper half of the stellar disk 
(Figs.~\ref{alfori2_model}b and \ref{alfori2_model}d).  
The CO gas within this patch is moving outward 
with a velocity of 5~\KMS, while the gas in the remaining region is moving 
inward faster with 25 \KMS.  
Figures~\ref{alfori2_model}a and \ref{alfori2_model}f show that 
the line profile and 
the wavelength dependence of the observed one-dimensional projection images 
within the 
CO line are reasonably reproduced, 
although the difference between the images in the blue wing and red wing 
is somewhat too pronounced, 
and the line profile is weaker than the observed data.  
We found out that the models with a patch of CO gas moving slowly outward at 
0--5~\KMS\ with the gas in the remaining region downdrafting 
much faster at 20--30~\KMS\ can reproduce the observed one-dimensional 
projection images reasonably. 

The above model can also explain why the MOLsphere in the blue wing is 
much more pronounced than in the red wing.  
Firstly, the velocity field with the weak upwelling and strong downdrafting 
components causes the 
line center to be redshifted with respect to the stellar rest frame. 
This can be seen in Fig.~\ref{alfori2_model}a, where 
the positions of the two transitions responsible for the observed line profile, 
(2,0) $R(26)$ and $R(75)$, are marked in the stellar rest frame.  
Since the contribution of the $R(26)$ transition to the line profile is much 
larger than that of $R(75)$ because of the much lower excitation potential 
of the former transition, the line profile is redshifted with respect to 
the position of $R(26)$ in the stellar rest frame.  
In other words, the wavelength of the stellar rest frame is located in the 
blue wing of the line.  
Secondly, the strong extended CO emission is seen in the line of 
sight tangential to the outer CO layer, because the column density along 
this line of sight is the largest.  The radial velocity of the CO 
layer along such a line of sight is nearly zero, which means that the 
strong extended CO emission appears at the stellar rest frame.  
Because the stellar rest frame is located in the blue wing as explained 
above, the extended CO emission is strong in the blue wing.  
The emission becomes nearly absent in the red wing, because the 
velocity difference between the blue and red wing is much greater than the 
line width.

The velocity field in 2009 is in contrast with that in 2008, which was 
characterized by the gas moving both outward 
and inward with velocities of 10--15~\KMS.  
Therefore, our AMBER observations at two epochs reveal a drastic change in 
the velocity field in the photosphere and MOLsphere within one year.

\section{Discussion}
\label{sect_discuss}

The drastic change in the velocity field between 2008 and 2009 sets an 
upper limit of one year on the time scale of the change of the MOLsphere.  
This allows us to estimate the upper limit of the radial spatial scale 
where the inhomogeneous gas motions are present.  
We assume that the upwelling patch (or clump) with 10--15~\KMS\ detected 
at 1.45~\RSTAR\ (radius of the MOLsphere) in 2008 decelerated linearly 
with time over one year and corresponds to the patch slowly moving outward 
with 0--5~\KMS\ in 2009.  
Then the maximum radial distance reached by the gas patch is 
0.24--0.47~\RSTAR.  
This means that the upwelling gas patch at 1.45~\RSTAR\ in 2008 can reach 
1.7--1.9~\RSTAR\ in 2009.  
Likewise, if we assume that the fast downdrafting patch with 20--30~\KMS\ 
detected in 2009 accelerated inward linearly with time starting 
from 0~\KMS, it must have traveled 0.71--1.07~\RSTAR\ in 1 year.  This suggests 
that the fast downdrafting gas patch could have been located as far as at 
2.2--2.5~\RSTAR\ in 2008 and could have fallen to 1.45~\RSTAR\ in 
one year.  
Therefore, the vigorous gas motions can be present up to $\sim$2~\RSTAR. 

These inhomogeneous gas motions in the extended atmosphere of Betelgeuse 
have also been detected by other observations.  
Recently Harper et al. (\cite{harper09a}) have studied the dynamics of the 
cool extended outer atmosphere of Betelgeuse based on high-spectral 
resolution mid-IR observations of the [Fe II] lines at 17.94 and 
24.53~\micron.   
These [Fe II] lines form at $\sim$1.6 \RSTAR\ 
(converted with the angular diameter of Betelgeuse derived here) in the cool 
extended outer atmosphere (see Fig.~8 of Harper et al. \cite{harper09a}) 
with estimated excitation temperatures of 1520--1950~K.  
Therefore, the [Fe II] lines originate in the region similar 
to the MOLsphere where the CO first overtone lines form.  
The profiles of the [Fe II] lines indicate turbulent gas motions without 
signatures of significant outflows of $\ga$10~\KMS. 
This is consistent with the velocity fields derived from our two-epoch AMBER 
observations. 
Harper et al. (\cite{harper09a}) detected no significant changes in 
the [Fe II] line profiles at three epochs over 14 months.  
However, this may be because the changes in the velocity field are 
smeared out in their spatially unresolved observations.  As can be seen in 
Fig. ~\ref{obsresCO}a, the CO line profiles observed with AMBER 
only show a low redshift, despite the remarkable change in the velocity 
field.  

Complex gas motions have been detected in the extended chromosphere 
of Betelgeuse as well.  
Lobel \& Dupree (\cite{lobel01}) present the modeling of the 
chromospheric velocity field up to $\sim$3~\RSTAR. 
Moreover, the 
velocity field changed from overall inward motions to outward motions 
within 0.5--1 year.  
Therefore, both the cool and hot components in the extended outer 
atmosphere are characterized by strongly temporally variable 
inhomogeneous gas motions.

The physical mechanism responsible for these vigorous motions and their 
drastic change within one year is not yet clear, although it is 
likely related to the unknown wind-driving mechanism.  
The convective energy flux is 
expected to be low in the MOLsphere, which extends to $\sim$1.3--1.4~\RSTAR. 
This poses a problem for the interpretation of the detected gas motions in 
terms of convection. 
Other possible mechanisms include Alfv\'en waves and pulsation.  
The recent detection of magnetic fields in Betelgeuse, albeit weak ($\sim$1 G), 
indicates that the prerequisite for Alfv\'en-driven-winds is available 
(Auri\`{e}re et al. \cite{auriere10}).  
Airapetian et al. (\cite{airapetian00}) show that Alfv\'en waves can 
drive mass outflows from the chromosphere 
with the velocity and mass-loss rate in agreement with those observed for 
Betelgeuse.   
However, 
the effects of the Alfv\'en-waves 
on the more dominant, cool outer atmosphere including the CO MOLsphere 
are not addressed.  
The MHD simulations of Suzuki (\cite{suzuki07}) for red giants show 
that the stellar winds are highly temporally variable and ``structured'', 
in which hot ($\ga \! 10^4$~K) gas bubbles are embedded in cool ($\sim$2000~K) 
gas (see, however, Airapetian et al. \cite{airapetian10} for critical 
discussion).  
The radial velocity within $\sim$10~\RSTAR\ also shows significant time 
variations from $\sim \! +40$~\KMS\ (outward motions) to $\sim \! -40$~\KMS\ 
(inward motions).  This is compatible to the change in the velocity field 
detected by our AMBER observations.  
However, the simulations of Suzuki (\cite{suzuki07}) were carried out 
for red giant stars, which are much less luminous ($\la \! 10^3$~\LSOL) 
compared to Betelgeuse ($1.3 \times 10^5$~\LSOL, 
Harper et al. \cite{harper08}).  
Extending the MHD simulations of Suzuki (\cite{suzuki07}) for more luminous 
stars, as well as the inclusion of the cool molecular component in the 
work of Airapetian et al. (\cite{airapetian00}), 
would be valuable for a comparison with the present and future 
AMBER observations.

Lobel (\cite{lobel10}) infers that strong shock waves generated by 
convection in the photosphere that are propagating outward 
may carry the energy and momentum to 
accelerate the wind and heat the chromosphere.  
The qualitative similarity between the inhomogeneous velocity field in the 
chromosphere and in the photosphere/MOLsphere may point toward this scenario.  
However, 
obviously, it is indispensable to map the dynamical structure of the 
cool outer atmosphere at various radii to clarify the driving 
mechanism of mass outflows in RSGs.

\section{Concluding remarks}
\label{sect_conclusions}

We have succeeded, for the first time, in one-dimensional aperture synthesis 
imaging of 
Betelgeuse in the individual CO first overtone lines, as well as in the 
continuum approximately free from molecular/atomic lines, with a spatial
resolution of 9.8 mas and a spectral resolution of 6000 using VLTI/AMBER.  

The one-dimensional projection images in the CO lines reconstructed with the 
self-calibration technique, which restores the complex visibility using 
differential phase measurements, reveal that the star appears different within 
the individual CO lines.   The one-dimensional projection images in the blue 
wing and line center show a pronounced extended component up to 1.3~\RSTAR, 
while the images in 
the red wing follow that in the continuum without an extended component.  
Our image reconstruction represents the 
first study to image the so-called MOLsphere of an RSG in the individual CO 
first overtone lines.  
Our modeling suggests that the dynamics in the photosphere and 
MOLsphere in 2009 is characterized by strong downdrafts with 20--30~\KMS\ and 
slight outward motions with 0--5~\KMS.  
This indicates a drastic change in the velocity field within one year from 
2008, when the dynamics was characterized by both upwelling and downdrafting 
components with 10--15~\KMS.  

On the other hand, the reconstructed one-dimensional projection images in the $K$-band 
continuum show only a small deviation of 5\% from the limb-darkened disk 
with an angular diameter of $42.49 \pm0.06$~mas with a power-law-type 
limb-darkening parameter of $(9.7 \pm 0.5) \times 10^{-2}$.  
This limb-darkened disk diameter results in an effective temperature of 
$3690 \pm 54$~K for the continuum-forming layer.  
The deviation from the limb-darkened disk in the one-dimensional projection 
images suggests that the amplitude of stellar spots is likely smaller than 
20--30\% of the intensity of the stellar disk.  
Furthermore, we detected no or only marginal time variation in the continuum 
visibility data within the measurement errors, 
much smaller than the maximum variation 
predicted by the current three-dimensional convection simulations. 
It cannot be entirely excluded that Betelgeuse showed unusually 
weak surface structures at the times of our AMBER observations just by 
chance. 
However, it is also possible that the current three-dimensional convection model for RSGs 
predicts too strong surface structures in the continuum.  
A long-term monitoring to measure the amplitude of the time variations 
in the visibility and closure phase is indispensable for a definitive, 
statistical test of three-dimensional convection simulations.  

The self-calibration imaging using differential phase has turned out to be 
very effective and necessary, 
despite the good linear $uv$ coverage from the first to fifth visibility lobe.  
This suggests that the self-calibration technique may be even more necessary 
for two-dimensional imaging, 
where it is difficult to obtain a $uv$ coverage as densely sampled as 
in our one-dimensional case.  
While the imaging of stellar surfaces 
is still challenging (e.g., Creech-Eakman et al. \cite{creech-eakman10}), 
our self-calibration one-dimensional imaging demonstrates 
a promising way to achieve that goal.

\begin{acknowledgement}
We thank the ESO VLTI team for supporting our AMBER observations.  
We are also grateful to Eric Thi\'ebaut, who kindly makes his image 
reconstruction software MiRA publicly available.  
NSO/Kitt Peak FTS data on the Earth's telluric features 
were produced by NSF/NOAO.
\end{acknowledgement}

\clearpage

\clearpage

\clearpage

\clearpage
\begin{figure*}
\sidecaption
\includegraphics[width=12cm]{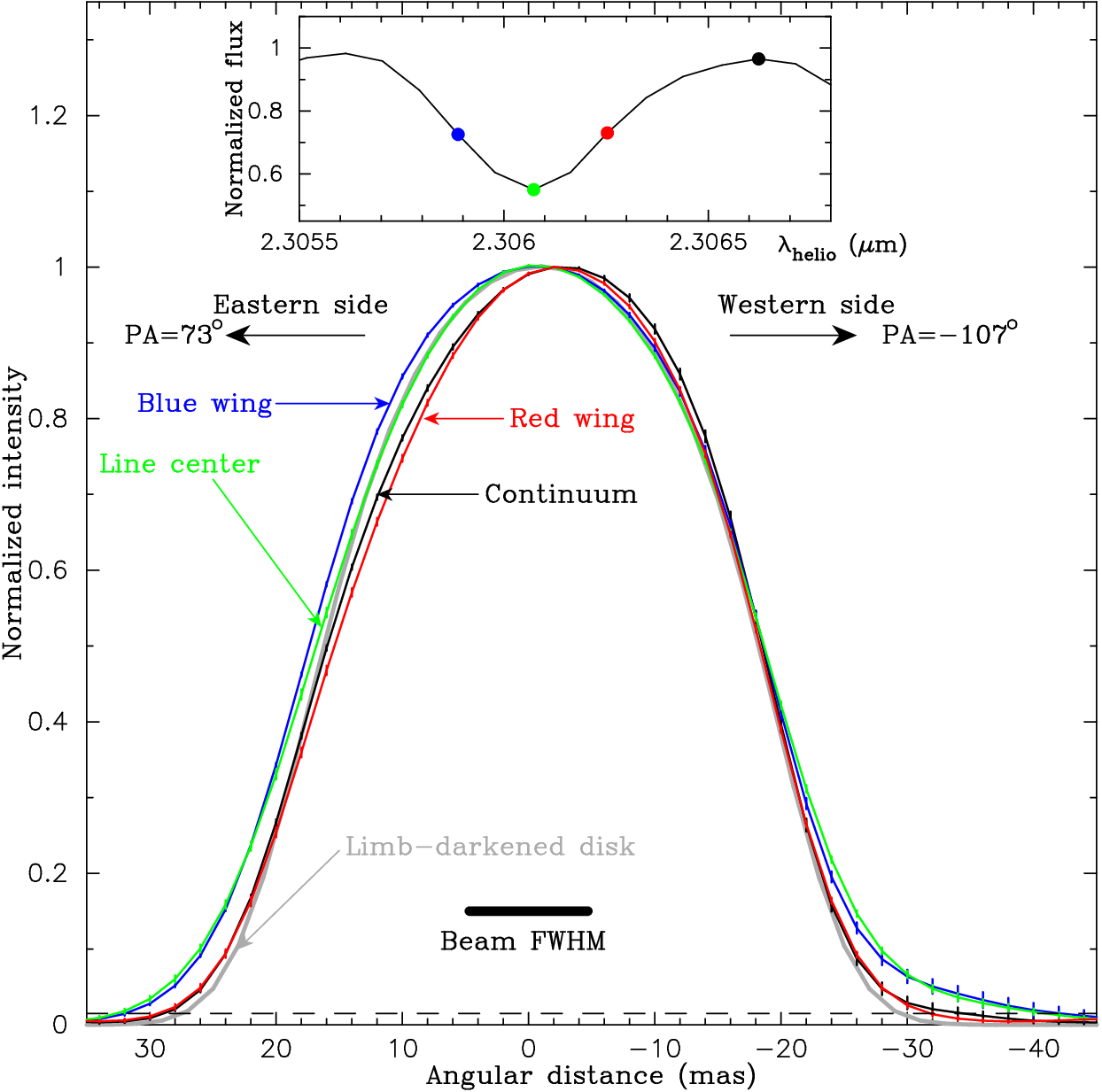}
\caption{
One-dimensional projection images reconstructed {\em at four different wavelengths 
(blue wing, line center, red wing, and continuum)} within the CO line due to the 
two transitions (2,0) $R(26)$ and $R(75)$ (spectral resolution of 6000) are 
shown by the blue, green, red, and black solid 
lines, respectively.  
The one-dimensional projection image of the limb-darkened disk with the parameters derived 
in Sect.~\ref{subsect_res_cont} is also shown by the gray solid line as a 
reference.  
The inset shows the observed line profile 
with the wavelengths of the images marked by the filled circles with 
the corresponding colors.  
The one-dimensional projection images are convolved with the Gaussian beam 
whose FWHM (9.8~mas) is shown by the thick solid line, and their 
absolute scale is normalized to the peak intensities.  
The orientation of the one-dimensional projection images is also shown.  
The dashed line denotes the noise level of the image reconstruction.  
}
\label{image_lines}
\end{figure*}

\clearpage
\begin{figure*}
\sidecaption
\includegraphics[width=12cm]{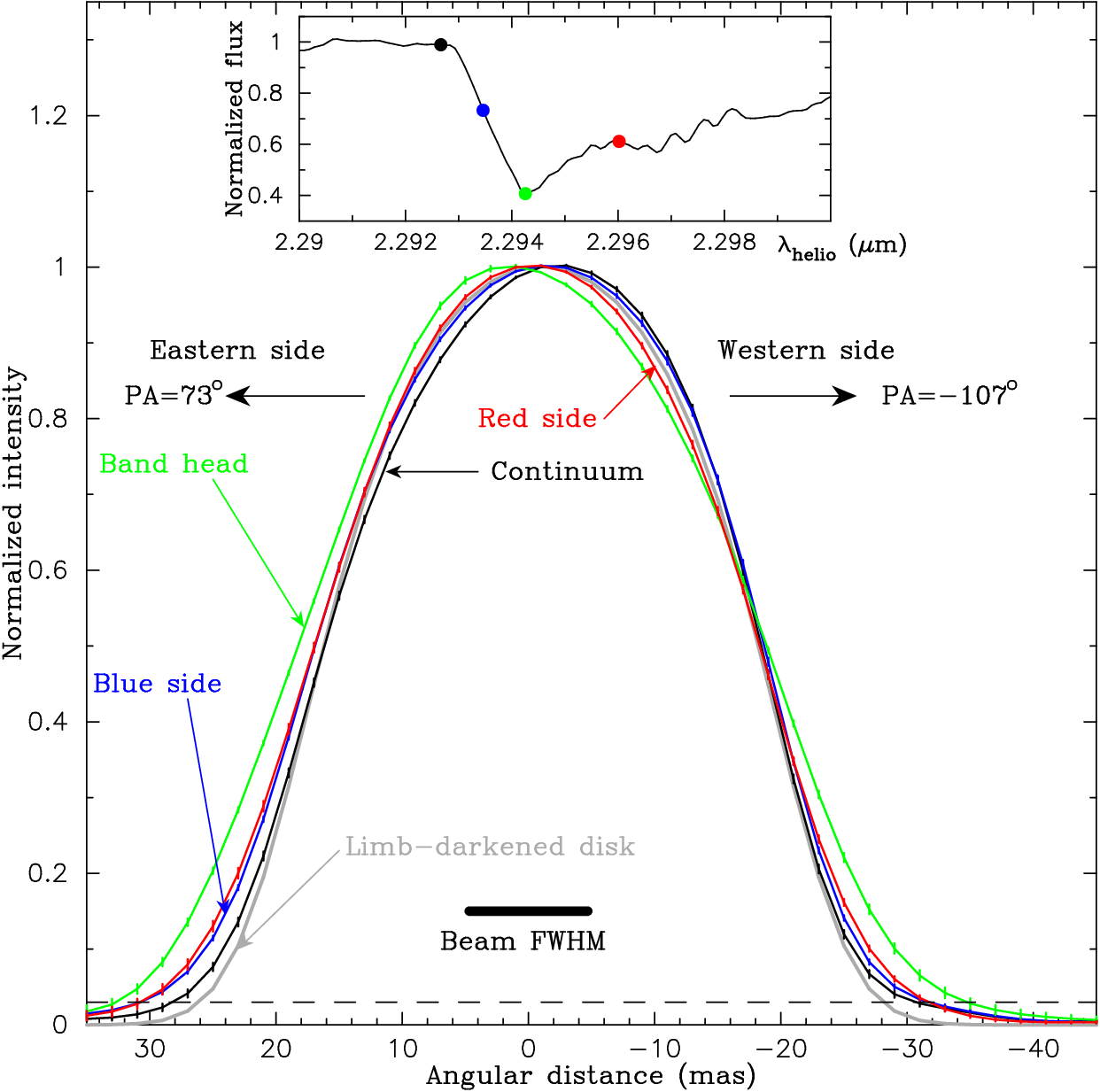}
\caption{
One-dimensional projection images reconstructed at four different wavelengths in the CO 
(2,0) band head (spectral resolution of 1600) shown in the same manner as 
Fig.~\ref{image_lines}.  
}
\label{image_bh}
\end{figure*}

\clearpage
\begin{figure*}
\resizebox{\hsize}{!}{\rotatebox{-90}{\includegraphics{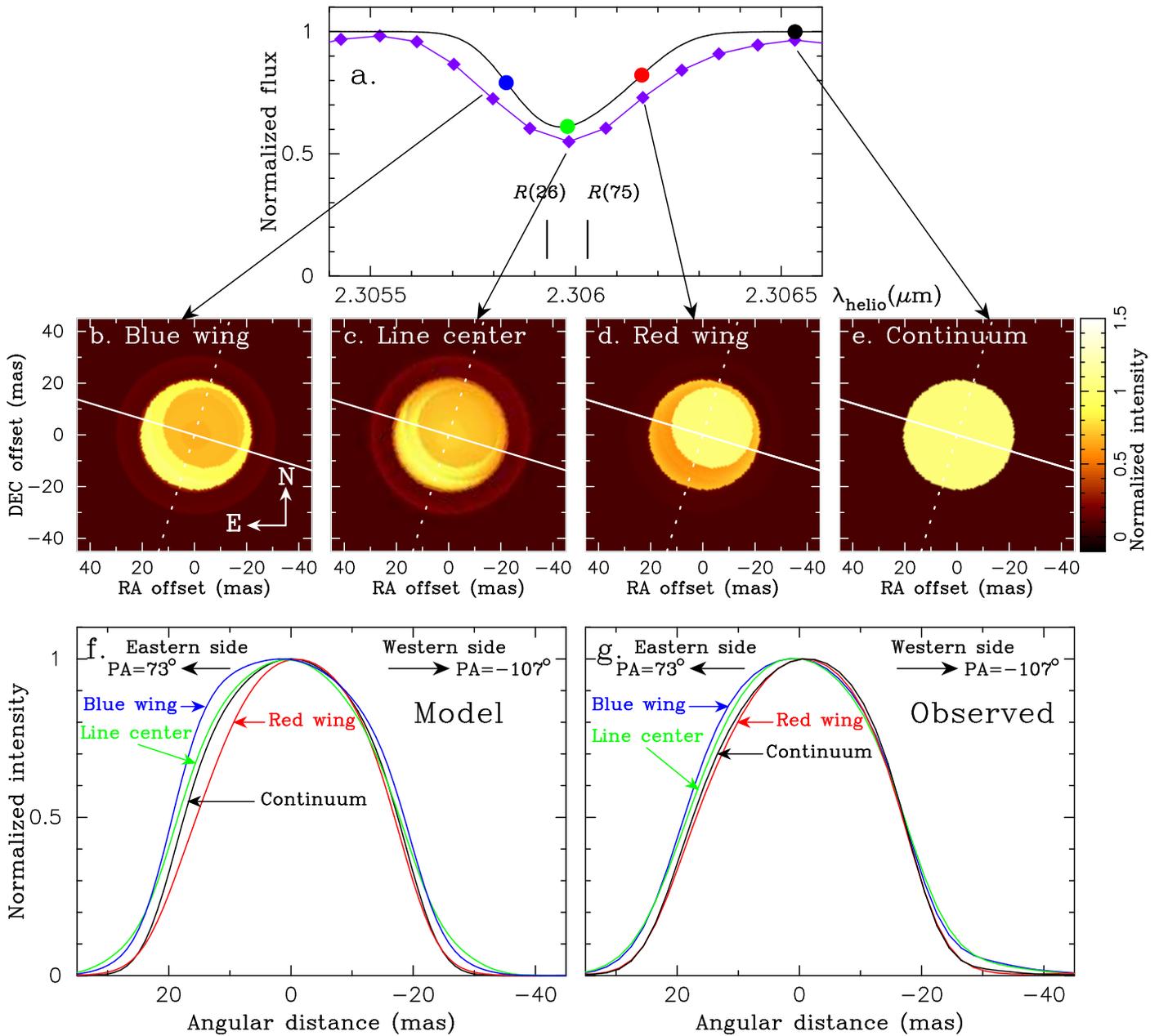}}}
\caption{
Best-fit stellar patch model with an inhomogeneous velocity field.  
{\bf a:} The filled diamonds represent the observed spectrum, while 
the black solid line represents the model spectrum.  The filled circles 
mark the wavelengths of the two-dimensional images and one-dimensional projection images shown 
in the panels {\bf b--g}. 
{\bf b--e:} Two-dimensional model images.  The orientation of the linear $uv$ 
coverage is shown by the solid lines.  The one-dimensional projection images 
are obtained by integrating the two-dimensional images in the direction shown by 
the dotted lines, which is perpendicular to the orientation of the linear 
$uv$ coverage. 
{\bf f:} The one-dimensional projection images predicted by the model at 
four wavelengths within the CO line.  
{\bf g:} The observed one-dimensional projection images for the same CO line. 
}
\label{alfori2_model}
\end{figure*}

\clearpage
\appendix
\section{Summary of AMBER observations}

Our AMBER observations of Betelgeuse and the calibrator Sirius are 
summarized in Table~\ref{obslog}. 

\begin{table*}
\begin{center}
\caption {Log of AMBER observations of Betelgeuse and the calibrator 
Sirius with the E0-G0-H0 (16-32-48m) baseline configuration.  Seeing is 
in the visible. 
}

\begin{tabular}{l l c c l r r | l l c c l r r}\hline
\# & $t_{\rm obs}$ & $B_{\rm p}$ & PA  & Seeing & $\tau_0$ & NDIT & 
\# & $t_{\rm obs}$ & $B_{\rm p}$ & PA  & Seeing  & $\tau_0$ & NDIT  \\ 
   &  (UTC)       &      (m)   &(\degr)&  (\arcsec)      & (ms)     &      & 
   &  (UTC)       &      (m)   &(\degr)&   (\arcsec)     & (ms)     &        \\
\hline
\multicolumn{14}{c}{Betelgeuse} \\
\hline
\multicolumn{7}{c|}{2009 Jan 5} & \multicolumn{7}{c}{2009 Jan 6} \\
\hline
1 &  01:01:44 & 22.24/11.12/33.36 & 73 & 0.76 & 6.2 &500  &24 &  00:47:33 & 21.24/10.63/31.87 & 73 & 1.55 & 3.8 &500\\
2 &  01:03:56 & 22.45/11.23/33.68 & 73 & 0.77 & 6.1 &500  &25 &  00:49:46 & 21.46/10.74/32.20 & 73 & 1.53 & 4.0 &500\\
3 &  01:06:09 & 22.66/11.33/33.99 & 74 & 0.77 & 6.1 &500  &26 &  00:51:59 & 21.68/10.84/32.52 & 73 & 1.42 & 4.7 &500\\
4 &  01:08:22 & 22.86/11.44/34.30 & 74 & 0.84 & 5.4 &500  &27 &  00:54:12 & 21.89/10.95/32.85 & 73 & 1.50 & 4.7 &500\\
5 &  01:10:35 & 23.07/11.54/34.61 & 74 & 0.89 & 5.3 &500  &28 &  00:56:25 & 22.11/11.06/33.17 & 73 & 1.41 & 5.0 &500\\
6 &  01:50:20 & 26.41/13.21/39.62 & 75 & 0.58 & 11.1 &500  &29 &  00:58:37 & 22.32/11.16/33.48 & 73 & 1.41 & 4.8 &500\\
7 &  01:52:34 & 26.57/13.29/39.87 & 75 & 0.58 & 11.2 &500  &30 &  01:48:34 & 26.57/13.29/39.86 & 75 & 1.09 & 8.4 &500\\
8 &  01:54:46 & 26.74/13.38/40.11 & 75 & 0.57 & 11.6 &500  &31 &  01:50:47 & 26.73/13.37/40.11 & 75 & 1.10 & 8.4 &500\\
9 &  01:56:59 & 26.90/13.46/40.36 & 75 & 0.58 & 12.2 &500  &32 &  01:53:00 & 26.89/13.45/40.35 & 75 & 1.01 & 9.4 &500\\
10 &   01:59:12 & 27.06/13.54/40.59 & 75 & 0.65 & 11.2&500   &33 &  01:55:13 & 27.05/13.53/40.59 & 75 & 1.04 & 9.5 &500\\
11 &   02:51:26 & 30.11/15.06/45.18 & 75 & 0.95 & 6.6 &500  &34 &  01:57:25 & 27.21/13.61/40.82 & 75 & 1.04 & 9.7 &500\\
12 &   02:53:39 & 30.21/15.11/45.32 & 75 & 0.88 & 7.2 &500  &35 &  02:32:55 & 29.40/14.71/44.10 & 75 & 1.27 & 16.4 &500\\
13 &   02:55:52 & 30.31/15.16/45.47 & 75 & 0.91 & 7.1 &500  &36 &  02:35:07 & 29.51/14.77/44.28 & 75 & 1.26 & 14.2 &500\\
14 &   02:58:04 & 30.40/15.21/45.61 & 75 & 1.02 & 6.3 &500  &37 &  02:37:20 & 29.63/14.82/44.45 & 75 & 1.30 & 11.8 &500\\
15 &   03:00:18 & 30.49/15.25/45.75 & 75 & 0.98 & 6.3 &500  &38 &  02:39:33 & 29.74/14.88/44.61 & 75 & 1.31 & 11.8 &500\\
16 &   04:07:52 & 31.98/16.00/47.98 & 73 & 1.19 & 5.0 &1000  &39 &  02:41:46 & 29.84/14.93/44.77 & 75 & 1.45 & 10.2 &500\\
17 &   04:12:08 & 31.99/16.01/47.99 & 73 & 1.12 & 5.3 &1000  &40 &  03:20:21 & 31.30/15.66/46.97 & 74 & 1.11 & 17.8 &500\\
18 &   04:16:24 & 31.99/16.00/47.99 & 73 & 1.09 & 5.6 &1000  &41 &  03:22:34 & 31.36/15.69/47.06 & 74 & 1.02 & 17.0 &500\\
19 &   05:00:17 & 31.41/15.72/47.13 & 71 & 0.89 & 12.4 &500  &42 &  03:24:47 & 31.42/15.72/47.14 & 74 & 0.99 & 16.3 &500\\
20 &   05:02:29 & 31.36/15.69/47.05 & 71 & 0.92 & 12.4 &500  &43 &  03:27:00 & 31.48/15.75/47.22 & 74 & 1.01 & 15.6 &500\\
21 &   05:04:42 & 31.30/15.66/46.96 & 71 & 0.94 & 12.1 &500  &44 &  03:29:13 & 31.53/15.77/47.30 & 74 & 0.90 & 16.0 &500\\
22 &   05:06:55 & 31.24/15.63/46.87 & 71 & 0.97 & 13.0 &500  &45 &  04:05:36 & 31.98/16.00/47.99 & 73 & 0.97 & --- &500\\
23 &   05:09:08 & 31.17/15.60/46.77 & 71 & 0.92 & 14.8 &500  &46 &  04:07:49 & 31.99/16.01/47.99 & 73 & 1.08 & --- &500\\
   &                 &                   &      &      &    &    &47 &  04:10:01 & 31.99/16.01/48.00 & 73 & 1.16 & --- &500\\
   &                 &                   &      &      &    &    &48 &  04:12:14 & 31.99/16.01/47.99 & 73 & 1.11 & --- &500\\
   &                 &                   &      &      &    &    &49 &  04:14:27 & 31.98/16.00/47.99 & 73 & 1.15 & --- &500\\
   &                 &                   &      &      &    &    &50 &  04:49:21 & 31.57/15.80/47.37 & 72 & 1.40 & --- &500\\
   &                 &                   &      &      &    &    &51 &  04:51:34 & 31.53/15.78/47.30 & 71 & 1.10 & --- &500\\
   &                 &                   &      &      &    &    &52 &  04:53:47 & 31.48/15.75/47.23 & 71 & 1.02 & --- &500\\
   &                 &                   &      &      &    &    &53 &  04:56:00 & 31.42/15.72/47.15 & 71 & 0.89 & --- &500\\
   &                 &                   &      &      &    &    &54 &  04:58:13 & 31.37/15.70/47.06 & 71 & 0.92 & --- &500\\
\hline

\multicolumn{14}{c}{Sirius} \\
\hline
\multicolumn{7}{c|}{2009 Jan. 5} & \multicolumn{7}{c}{2009 Jan. 6} \\
\hline
C1 & 01:28:18 & 24.75/12.38/37.13 & 50 & 0.84 & 5.9 & 2500 & C5 & 00:24:19 &
20.99/10.50/31.50 & 36 & 1.32 & 3.9 & 2000 \\
C2 & 02:17:35 & 27.84/13.93/41.76 & 59 & 0.96 & 8.5 & 5000 & C6 & 01:27:00 & 
24.92/12.47/37.38 & 51 & 1.34 & 5.9 & 2500 \\
C3 & 03:23:41 & 30.83/15.43/46.26 & 66 & 1.04 & 6.0 & 3000 & C7 & 02:09:58 & 
27.57/13.79/41.36 & 58 & 1.02 & 12.9 & 2500  \\
C4 & 04:36:43 & 31.99/16.01/47.99 & 73 & 0.90 & 7.5 & 2500 & C8 & 02:56:21 & 
29.91/14.97/44.88 & 64 & 2.01 & 54.0 & 2500 \\
   &          &                   &    &    &     &        & C9 & 03:43:15 & 
31.46/15.74/47.20 & 69 & 0.81 & 22.4 & 2500 \\
   &          &                   &    &    &      &        &C10 & 04:26:45 & 
31.99/16.01/47.99 & 72 & 1.56 & 41.2 & 2500 \\
\hline

\label{obslog}

\end{tabular}
\end{center}
\end{table*}

\section{Image reconstruction of simulated data}
\label{appendix_simtests}

\begin{figure*}
\sidecaption
\includegraphics[width=12cm]{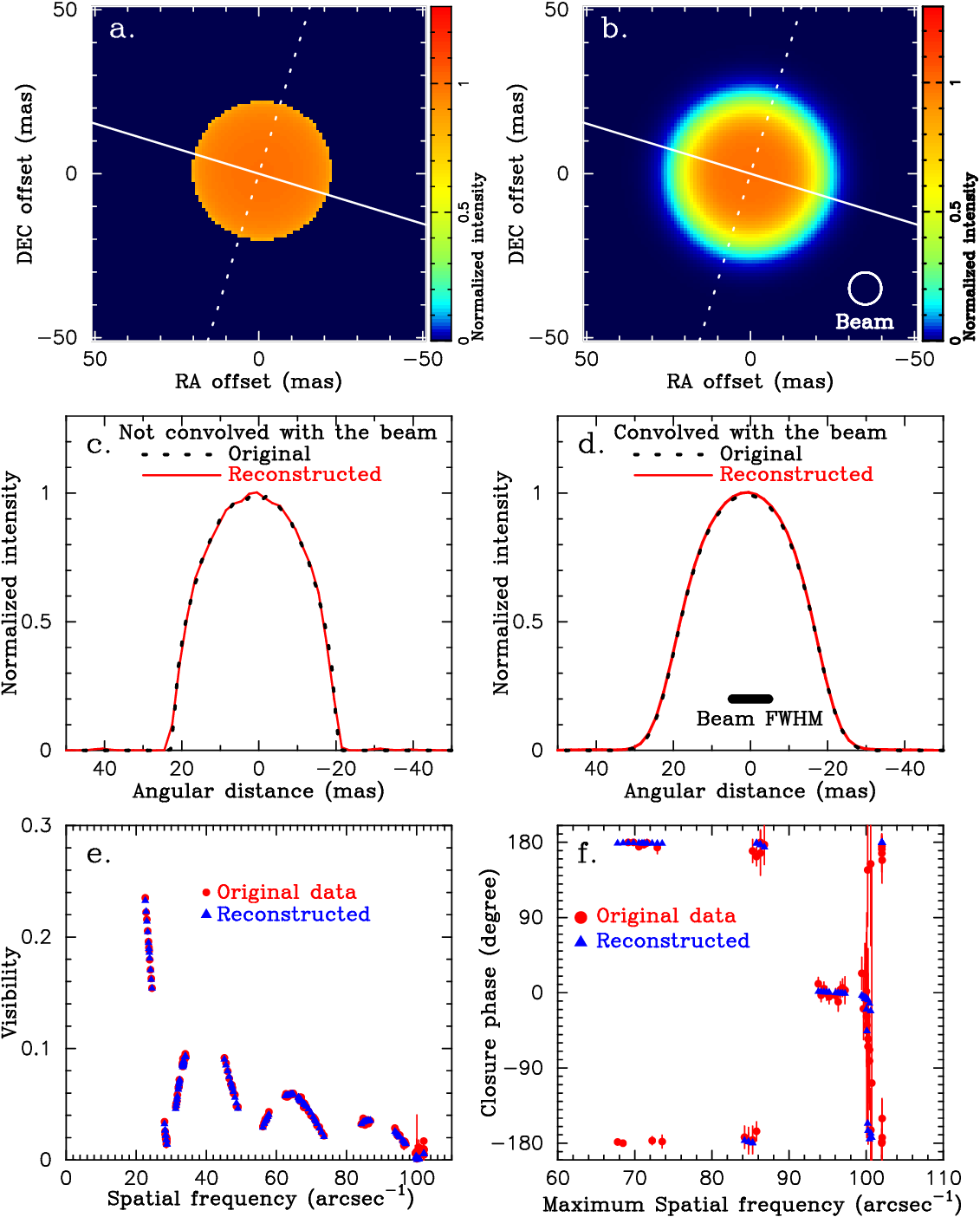}
\caption{
Image reconstruction of the simulated data for a limb-darkened disk with the 
parameters derived in Sect.~\ref{subsect_res_cont}. 
{\bf a:} Original two-dimensional image of the simulated data.  The solid line represents 
the orientation of the linear $uv$ coverage, while the dotted line 
represents the orientation perpendicular to it.  
{\bf b:} Two-dimensional image of the simulated data convolved with the Gaussian beam with 
a FWHM of 9.8~mas.  
{\bf c:} Comparison between the original and reconstructed one-dimensional projection 
images before convolving with the Gaussian beam.  The one-dimensional projection images 
are obtained by integrating the two-dimensional images in the direction shown 
by the dotted lines in the panels {\bf a} and {\bf b}.  
{\bf d:} Comparison between the original and reconstructed one-dimensional projection 
images convolved with the Gaussian beam with a FWHM of 9.8~mas. 
{\bf e:} The filled circles and triangles represent the visibilities from the 
original simulated data and the reconstructed image, respectively.  
{\bf f:} The filled circles and triangles represent the CPs from the 
original simulated data and the reconstructed image, respectively.  
The abscissa is the spatial 
frequency of the longest baseline of each data set.  
}
\label{simdata_ldd}
\end{figure*}

\begin{figure*}
\sidecaption
\includegraphics[width=12cm]{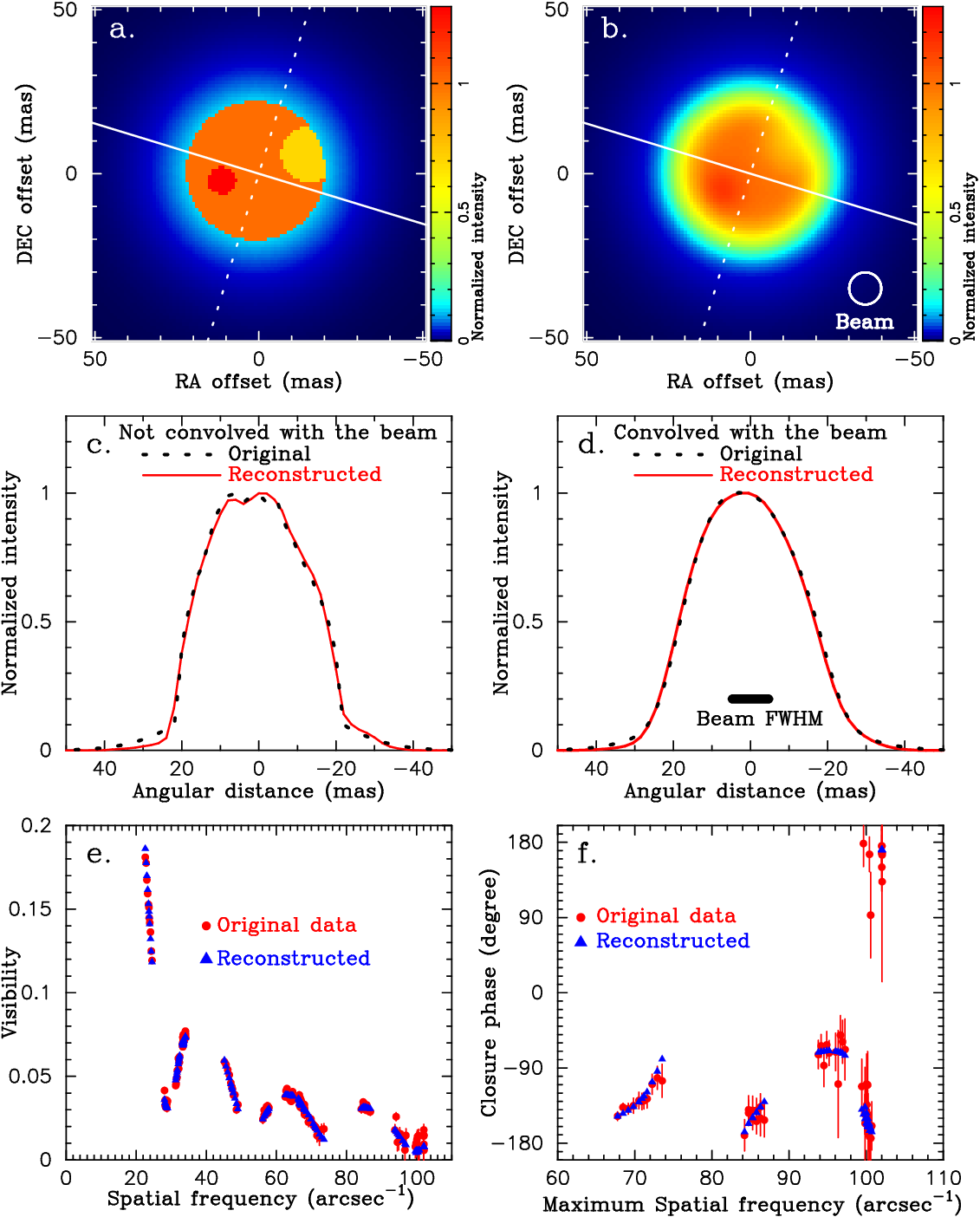}
\caption{
Image reconstruction of the simulated data for a uniform disk with a bright 
spot, a dark spot, and a halo shown in the same manner as in 
Fig.~\ref{simdata_ldd}.  
}
\label{simdata_udspots}
\end{figure*}

Aperture synthesis imaging from optical/IR interferometric data depends 
on a number of parameters used in the image reconstruction process, such 
as the initial model, as well as the regularization scheme and prior, 
which represent {\em a priori} information about the object's intensity 
distribution.  
Therefore, it is important to carry out the image reconstruction 
for simulated images to derive the appropriate reconstruction parameters 
before we attempt the image reconstruction from observed data.  
From simulated images, we generate simulated interferometric data by 
sampling the visibilities and CPs at the same $uv$ points 
as interferometric observations.  
With the true images known for these simulated data, we can examine 
the appropriate reconstruction parameters that allow us to reconstruct 
the original images correctly.  

Because MiRA is developed for two-dimensional image reconstruction, we took the 
following approach for the reconstruction of one-dimensional projection 
images: two-dimensional image reconstruction was carried out using 
an appropriate initial model and regularization parameters as described 
below.  
The reconstructed two-dimensional image was convolved with the clean beam,
which is represented by a two-dimensional Gaussian with a FWHM of 
$\lambda/B_{\rm max}$ = 9.8 mas, 
where $B_{\rm max}$ is the maximum baseline length of the data.  
The one-dimensional projection image was obtained by integrating 
this convolved two-dimensional image in the direction perpendicular to 
the linear $uv$ coverage.

We generated two simulated images that represent possible surface patterns 
of Betelgeuse: a simple limb-darkened disk and a uniform disk with a bright 
spot, a dark spot, and an extended halo, 
as shown in Figs. \ref{simdata_ldd}a and \ref{simdata_udspots}a, 
respectively. 
For both cases, the stellar angular diameter 
was set to be 42.5~mas, which is the limb-darkened disk diameter derived from 
all continuum visibilities measured in 2009.  
The visibilities and CPs were computed from the simulated images at the same 
$uv$ points as our AMBER observations, using the program of one of the 
authors (K.-H. Hofmann).  Noise was also added to the simulated visibilities 
and CPs to achieve SNRs similar to the AMBER data.  
We tested different initial models, priors, and regularization schemes to 
find out the appropriate parameter range 
to reconstruct the one-dimensional projection image of 
the simulated data correctly.  
It turned out that uniform disks with angular diameters between 34 and 50~mas 
serve as good initial models.  The prior used in the present work is a 
smoothed uniform disk described as 
\[
Pr(r) = \frac{1}{e^{(r-r_{\rm p})/\varepsilon_{\rm p}} + 1},
\]
where $r$ is the radial coordinate in mas, and $r_{\rm p}$ and $\varepsilon_{\rm p}$ 
define the size and the smoothness of the edge ($\varepsilon_{\rm p}
\rightarrow 0$ corresponds to a uniform disk), respectively.   
The appropriate values for $r_{\rm p}$ and $\varepsilon_{\rm p}$ were found to
be 10 .. 15 (mas) and 2 .. 3 (mas), respectively.  
Therefore, we used six different parameter sets 
for the image reconstruction of Betelgeuse 
by combining three diameters for the
initial uniform-disk model (34, 42, and 50 mas) and two different parameter 
sets for the prior (($r_{\rm p}, \varepsilon_{\rm p}$ = (10, 2) and (15, 3)).  
The final images and their uncertainties were obtained by taking the 
average and standard deviation, respectively, from the results 
reconstructed with these six parameter sets.  
The regularization using the maximum entropy method turned out to be 
appropriate for our reconstruction.  
We started the reconstruction with a high degree of regularization 
($\mu = 10^5$, see Thi\'ebaut \cite{thiebaut08} for the definition of 
$\mu$) and reduced it gradually by a factor of 10 after every 
500 iterations until the reduced $\chi^2$ reaches $\sim$1 or MiRA 
stops the iteration.  
These tests with 
the simulated data also confirm the validity of our approach to reconstruct 
one-dimensional projection images using the MiRA software for two-dimensional 
image reconstruction.

\section{Self-calibration imaging with differential phase}
\label{appendix_selfcal}

Because the principle of the self-calibration technique using DP 
measurements is described in detail in Millour et al. 
(\cite{millour11}), we mention the actual procedure only briefly.  
Then we describe the modification we added to this technique 
to deal with an issue specific to the AMBER data of Betelgeuse.  

The DP measured with AMBER at each $uv$ point 
contains information on the phase of the complex visibility function 
and roughly represents the difference between the phase in a spectral 
feature and that in the continuum.  
However, two pieces of information are lost because of the atmospheric 
turbulence: the absolute phase offset and the linear phase gradient with 
respect to wavenumber.  
We can derive this lost phase offset and gradient by a linear 
fit to the phase (as a function of wavenumber) from the reconstructed 
continuum images, if the image reconstruction in the 
continuum is reliable and not sensitive to the reconstruction parameters.  
This is indeed the case for our image reconstruction of Betelgeuse in 
the continuum, as discussed in Sect~\ref{subsect_img_cont}.  
Therefore, the phase in the CO lines can be restored from the continuum 
phase interpolated at the line spectral channels and the DPs 
measured in the lines.  
The image reconstruction is carried out with the measured visibilities 
and CPs as well as the restored phase.  
This process can be iterated, but our experiments show that the 
reconstructed images do not change after the first iteration.

We added the following modification to the technique presented in 
Millour et al. (\cite{millour11}).  
When the phase offset and gradient are derived by a linear fit to the 
phase of the reconstructed images, 
we only use the continuum spectral channels below 2.293~\micron\ and 
those between the adjacent CO lines above 2.3~\micron, instead of 
using the entire spectral channels, as Millour et al. (\cite{millour11}) 
did.  
The reason for this selection of the spectral channels is that 
the image reconstruction near the CO band head at 2.294~\micron\ is so 
uncertain owing to the poor SNR in the data binned with a spectral 
resolution of 6000 that 
the inclusion of the spectral channels near the band head in the linear 
fit hampers the reliable derivation of the phase offset and gradient.  

\begin{figure}
\resizebox{\hsize}{!}{\rotatebox{0}{\includegraphics{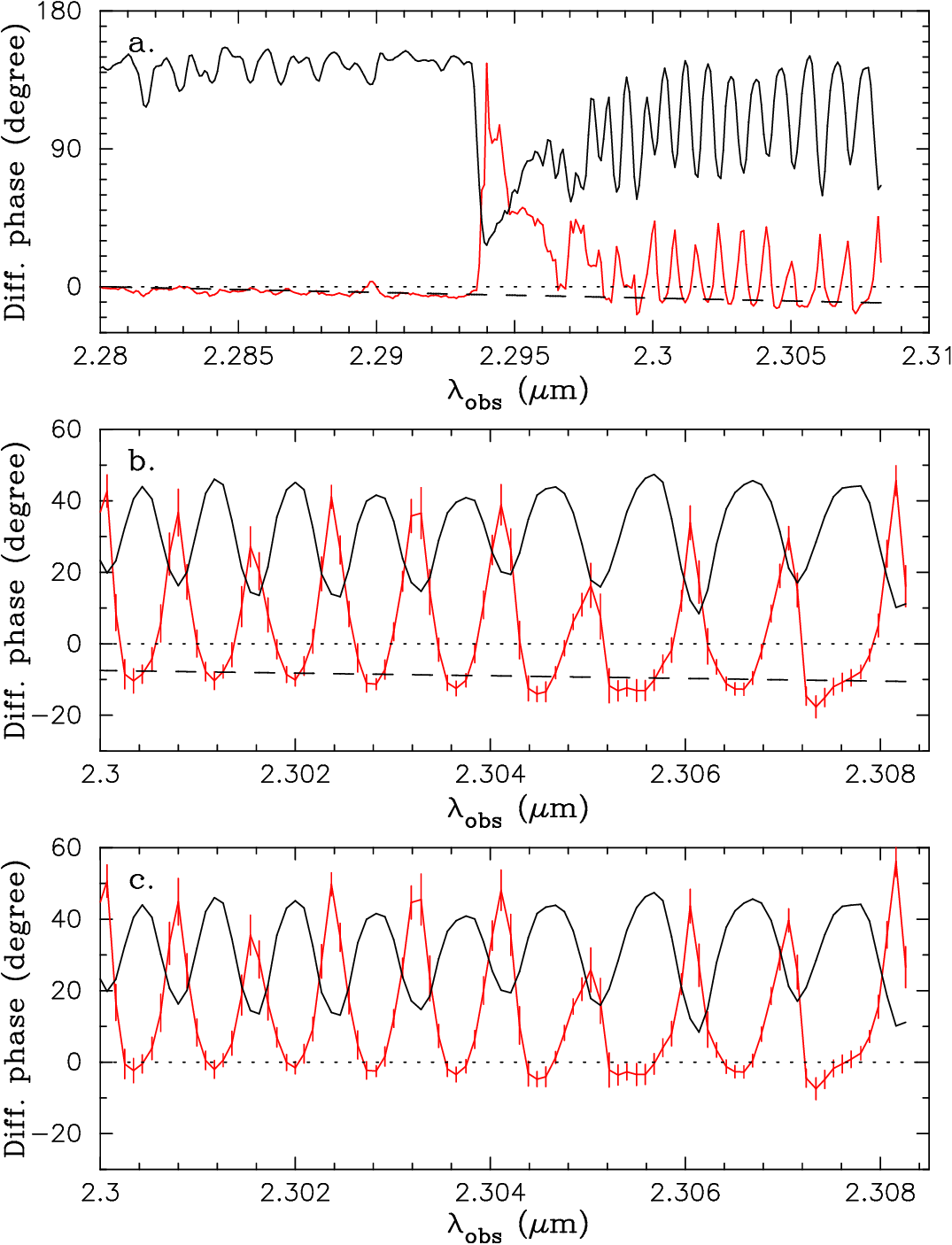}}}
\caption{
{\bf a:} DP observed on the longest baseline in the 
data set \#9 ($B_{\rm p}$ = 40.36~m) is plotted by the red solid line. 
The black line represents the scaled 
observed spectrum.  The DP and spectrum are binned with a spectral 
resolution of 6000.  
The dashed line represents the linear fit at the selected continuum 
points as described in Appendix~\ref{appendix_selfcal}.  
DP = 0 is shown by the dotted line.  
{\bf b:} Enlarged view of the panel {\bf a} for the CO lines. 
Note that the DP in the continuum points between 
the adjacent CO lines deviates from zero.  
{\bf c:} DP after subtracting the linear fit to the continuum points 
as described in Appendix~\ref{appendix_selfcal} is shown by the 
red solid line.  The black line represents the scaled spectrum.   
The DP in the continuum points between the adjacent CO lines is now zero 
within the measurement errors. 
}
\label{refitDP}
\end{figure}

The inclusion of only the selected continuum channels has the following 
consequence. 
If we denote the continuum phase from the reconstructed continuum image
at a given baseline and at the $i$-th spectral 
channel as $\varphi_{\rm c}(i)$, the phase at the $i$-th spectral channel, 
$\varphi (i)$, is restored as 
\[
\varphi (i) = \varphi_{\rm c}(i) + {\rm DP}(i), 
\]
where ${\rm DP}(i)$ represents the differential phase at the 
$i$-th spectral channel measured at the same baseline.  
At a continuum spectral channel denoted as $i_{c}$, the restored phase 
$\varphi (i_{\rm c})$ should be equal to the phase from the reconstructed 
continuum image $\varphi_{\rm c}(i_{\rm c})$.  This is fulfilled if the 
measured DP in the continuum is zero.  
However, the measured DPs show noticeable non-zero values in the continuum, 
as exemplarily shown in Figs.~\ref{refitDP}a and \ref{refitDP}b.  
The reason for the non-zero DPs in the continuum is that 
amdlib derives differential phase by a linear fit to 
the instantaneous phase at {\em all} spectral channels.  
Owing to the strong deviation of the phase in many CO lines from that in the 
continuum, this linear fit does not go through all the 
continuum points.  
Therefore, the non-zero DPs in the continuum spectral channels lead to 
a systematic error in the phase restored in the continuum, which 
affects the subsequent image reconstruction.  
We found out that the continuum one-dimensional projection image reconstructed 
using the restored phase shows a systematic wavelength dependence from the 
shortest to the longest wavelength of the observed spectral range, which is 
not seen in the continuum images reconstructed from the visibilities and CPs 
alone.

It is necessary to use the same spectral channels in the linear fit to 
the phase for the derivation of DP and for the derivation of the phase offset 
and gradient.  
Therefore, we refitted the DP from amdlib with a linear 
function (with respect to wavenumber) at the same continuum points 
as used for the derivation of the phase offset and gradient 
(dashed lines in Figs.~\ref{refitDP}a and \ref{refitDP}b) 
and subtracted the fitted linear function from the observed DP.  
This procedure enforces the DP in the continuum spectral channels 
to zero within the measurement errors, as 
shown in Fig.~\ref{refitDP}c.  
The phase was restored using this ``refitted'' DP.  
The continuum one-dimensional projection images reconstructed using the refitted 
DPs do not show the aforementioned systematic wavelength dependence, 
which proves the validity of our procedure. 

\clearpage

\section{Fit to the interferometric data with the reconstructed images}
\label{appendix_fit}

\begin{figure*}
\resizebox{\hsize}{!}{\rotatebox{-90}{\includegraphics{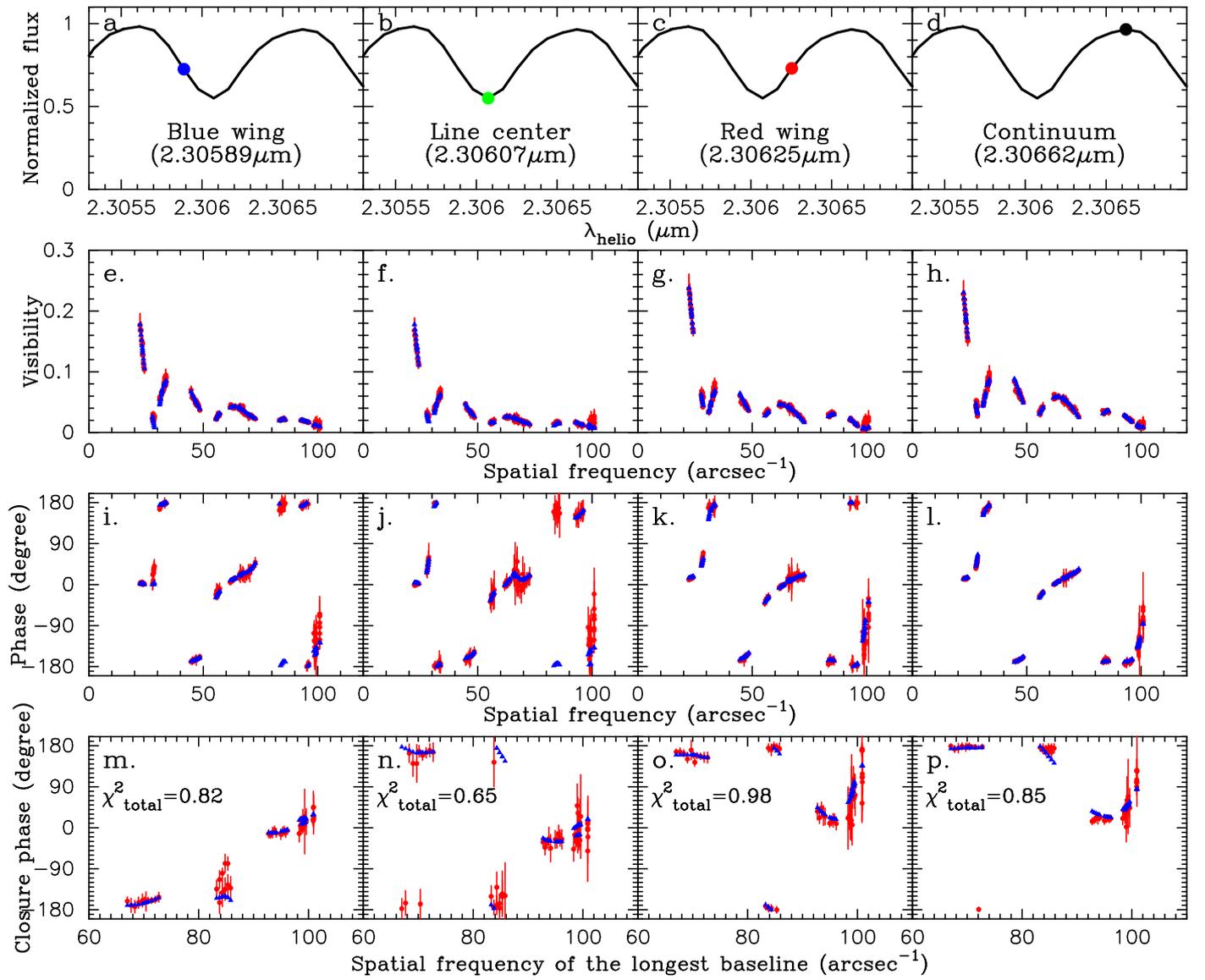}}}
\caption{
Comparison between the observed interferometric data and those 
from the one-dimensional projection image reconstruction for the CO line shown in 
Fig.~\ref{image_lines}.  
The first, second, third, and fourth columns show the comparison 
for the blue wing, line center, red wing, and continuum, respectively.  
The panels in the top row ({\bf a}--{\bf d}) show the observed CO line 
profile, and the filled circles denote the wavelength of the data 
shown in each column.  
In the remaining panels, the observed data are 
represented by the red circles, while the values from the image 
reconstruction are shown by the blue triangles.  
The reduced $\chi^2$ values for the fit including the complex visibilities, 
squared visibilities, and CPs are given in the panels {\bf m}--{\bf p}.  
}
\label{fit_selfcal}
\end{figure*}

\begin{figure*}
\resizebox{\hsize}{!}{\rotatebox{-90}{\includegraphics{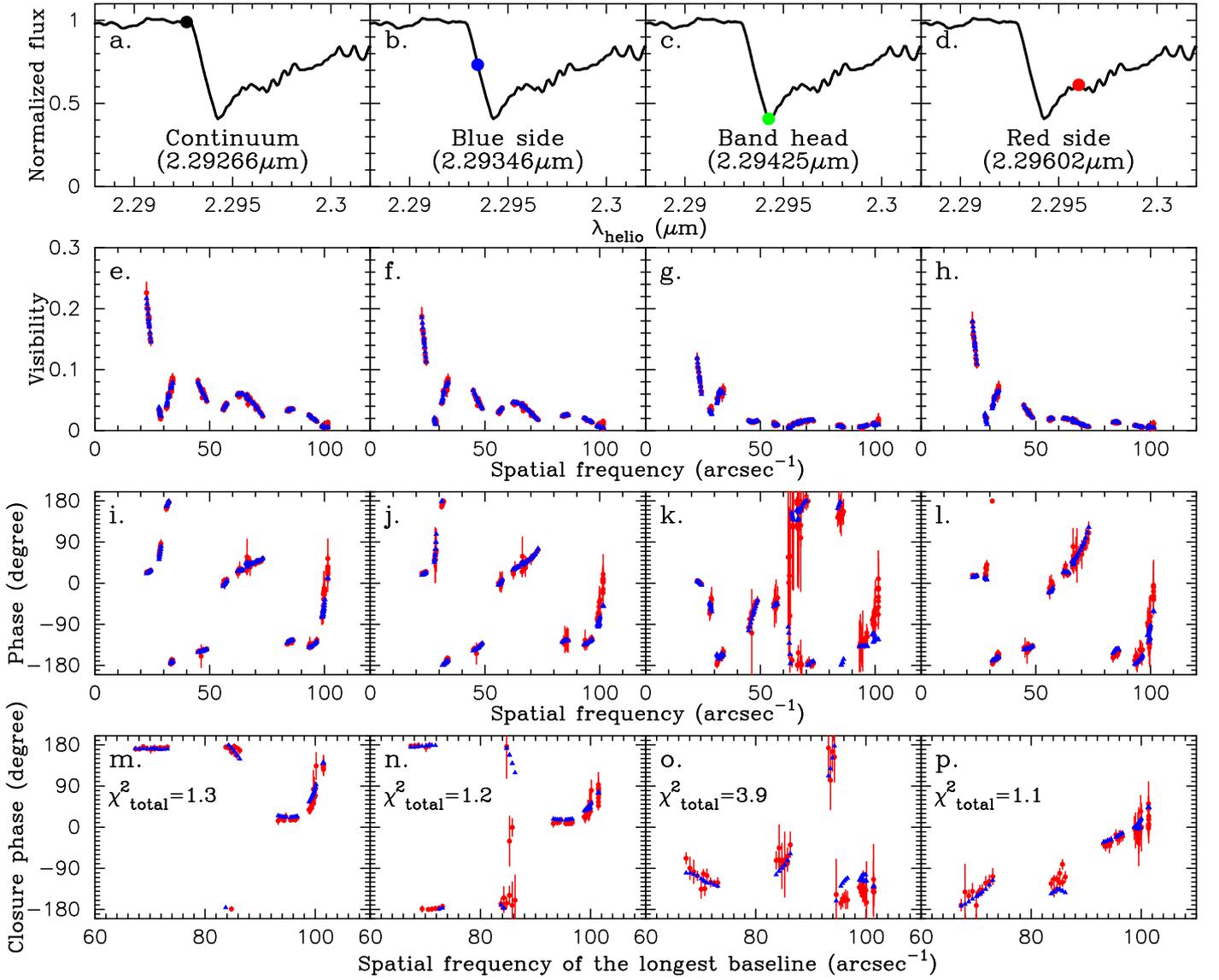}}}
\caption{
Comparison between the observed interferometric data and those 
from the one-dimensional projection image reconstruction near the CO (2,0) band head shown in 
Fig.~\ref{image_bh}.  
The first, second, third, and fourth columns show the comparison 
for the continuum, blue side between the continuum and the band head, 
bottom of the band head, and red side of the band head, 
respectively.  
The panels in the top row ({\bf a}--{\bf d}) show the observed spectrum of 
the CO band head, and the filled circles denote the wavelength of the data 
shown in each column.  
In the remaining panels, the observed data are 
represented by the red circles, while the values from the image 
reconstruction are shown by the blue triangles.  
The reduced $\chi^2$ values for the fit including the complex visibilities, 
squared visibilities, and CPs are given in the panels {\bf m}--{\bf p}.  
}
\label{fit_bh}
\end{figure*}

Figures~\ref{fit_selfcal} and \ref{fit_bh} 
show the fit to the observed interferometric data for the one-dimensional projection image 
reconstruction in the CO line and in the CO 
(2,0) band head shown in Figs.~\ref{image_lines} and \ref{image_bh}, 
respectively.  

\end{document}